\documentclass[fleqn,10pt]{wlscirep}
\usepackage[utf8]{inputenc}
\usepackage[T1]{fontenc}
\usepackage{bm}
\usepackage{graphicx}%
\usepackage{multirow}%
\usepackage{amsmath,amssymb,amsfonts}%
\usepackage{amsthm}%
\usepackage{mathrsfs}%
\usepackage[title]{appendix}%
\usepackage{xcolor}%
\usepackage{textcomp}%
\usepackage{manyfoot}%
\usepackage{booktabs}%
\usepackage{algorithm}%
\usepackage{algorithmicx}%
\usepackage{algpseudocode}%
\usepackage{listings}%
\usepackage{physics}
\usepackage{soul}
\usepackage{gensymb} 
\usepackage{units}

\DeclareMathSymbol{\shortminus}{\mathbin}{AMSa}{"39}

\title{Relativistic and nonrelativistic spin splitting above and below the Fermi level in a \emph{g}-wave altermagnet}

\author[$\star$,1,2,$\dagger$]{Nicholas Dale}

\author[$\star$,1,2,3]{Omar A. Ashour}

\author[1,2]{Marc Vila}

\author[5,6]{Resham B. Regmi}

\author[1,2,4]{Justin Fox}

\author[2]{Cameron W. Johnson}

\author[2]{Edward S. Barnard}

\author[7]{Alexei Fedorov}

\author[2]{Alexander Stibor}

\author[5,6]{Nirmal J. Ghimire}

\author[1,2,$\S$]{Sinéad M. Griffin}

\affil[1]{Materials Sciences Division, Lawrence Berkeley National Lab, 1 Cyclotron Road, Berkeley, CA 94720, USA}

\affil[2]{Molecular Foundry, Lawrence Berkeley National Lab, 1 Cyclotron Road, Berkeley, CA 94720, USA}

\affil[3]{Department of Physics, University of California, Berkeley, CA 94720, USA}

\affil[4]{Department of Physics, Carnegie Mellon University, Pittsburgh, PA 15213, USA}

\affil[5]{Department of Physics and Astronomy, University of Notre Dame, Notre Dame, IN 46556, USA}

\affil[6]{Stravropoulos Center for Complex Quantum Matter, University of Notre Dame, Notre Dame, IN 46556, USA}

\affil[7]{Advanced Light Source, Lawrence Berkeley National Lab, 1 Cyclotron Road, Berkeley, CA 94720, USA}

\affil[$\star$]{Authors contributed equally to this work.}
\affil[$\dagger$]{e-mail: ndale@lbl.gov}
\affil[$\S$]{email: sgriffin@lbl.gov}

\begin{abstract}
Nonrelativistic spin splitting (NRSS) challenges conventional wisdom about antiferromagnets by allowing spin-split electronic bands even in collinear orders with zero net magnetization. This sub-class of antiferromagnets, recently dubbed "altermagnets,” enforces distinctive spin textures via spin-group symmetries in the crystal. However, direct experimental evidence for such symmetry-driven magnetism remains scarce, and distinguishing it from relativistic spin splitting presents additional challenges. Here, we combine first-principles calculations, symmetry analysis, and two spin-resolved spectroscopies—angle-resolved photoemission (spin-ARPES) and our newly developed spin- and angle-resolved electron reflection spectroscopy (spin-ARRES)—to achieve the first complete momentum-resolved mapping of relativistic (RSS) and nonrelastivistic (NRSS) spin splitting in CoNb$_4$Se$_8$. By probing both the occupied (spin-ARPES) and unoccupied (spin-ARRES) electronic states in a single experiment, we uncover a series of momentum-dependent spin splitting phenomena each of which switch sign under sixfold rotations and persists far above and below the Fermi level. Crucially, distinct nodal planes in momentum space distinguish NRSS from RSS features. Additionally, the observed collapse of NRSS and the persistence of RSS above the Néel temperature, distinguishes a genuine magnetic phase transition from inversion symmetry breaking. Our work demonstrates, for the first time, the combined power of spin-ARPES and spin-ARRES in capturing the full spin texture across an extended energy range, positioning CoNb$_4$Se$_8$ as a prototype for exploring spin-group-based phenomena. These findings open new routes for engineering spin-based functionalities ranging from neuromorphic computing to unconventional superconductivity in layered antiferromagnets. 
\end{abstract}
\begin{document}

\flushbottom
\maketitle

\thispagestyle{empty}

\section*{Introduction}
The resurgence of interest in spin-split antiferromagnets (AFMs)~\cite{Noda_et_al:2016, Smejkal2020, Naka2019, Ahn_et_al:2019, Hayami2019, Yuan_et_al:2020}, first suggested in the 1960s~\cite{pekar_combined_1964}, has driven a renewed effort to understand their unique magnetic and electronic order. Recent advances have provided a clearer framework for these materials, including order parameter descriptions~\cite{Hayami2018_multipole, Hayami2019, Hayami2020} and group-theoretical classifications~\cite{Smejkal2022a}. Unlike conventional AFMs, which exhibit doubly-degenerate electronic bands across the entire Brillouin zone (BZ), these so-called `altermagnets'~\cite{Smejkal2022} feature momentum-dependent spin splitting arising from intrinsic symmetries rather than relativistic effects. This non-relativistic spin splitting (NRSS) arises from the rich symmetries of spin groups~\cite{brinkmanSpaceGroupTheory1966, brinkmanTheorySpinspaceGroups1996, litvinSpinGroups1974, litvinSpinPointGroups1977, liuSpinGroupSymmetryMagnetic2022, Smejkal2022}, which completely decouple spin and real space in the absence of spin-orbit coupling. In NRSS materials, the charge densities of the two oppositely-polarized spin sublattices map onto each other through space group operations that are neither a translation nor an inversion, such as glide and screw operations. Within the BZ, the band structure is only spin degenerate at a set of nodal surfaces, and spin-split elsewhere. The number of the nodal surfaces leads to specific magnetic states such as the experimentally realized \emph{d}-wave (2 nodal surfaces)~\cite{Jiang2025_inverselieb} and \emph{g}-wave (4 nodal surfaces)~\cite{Krempasky_et_al:2024, Reimers_et_al:2024} phases.  The understanding and observation of NRSS unlocks functional properties of both fundamental and technological interest ranging from unconventional superconductivity~\cite{Mazin:2022,decarvalho2024_am_sc_soc,chakraborty2024_am_sc} and novel topological effects~\cite{das2024_nem_topo_am,li2024_topo_weyl_am} to energy-efficient microelectronics\cite{jungwirth_2025_altm_spintron,Duan2025_altm_neuromorph}.


However, experimentally observing NRSS in materials remains a formidable challenge due to its subtle and often indirect nature. While theory predicts a rich variety of spin-split states stemming from symmetry-enforced mechanisms, translating these predictions into direct experimental observations is nontrivial and faces several key obstacles.

First and foremost, the fabrication of high-quality samples poses a significant barrier. Many candidate materials suffer from domain formation~\cite{Hariki2024, Amin2024}, competing ground states~\cite{smolyanuk2024}, mosaicity in thin films~\cite{Fields_2024_mosaicity}, or degradation during transport between growth and measurement chambers~\cite{Reimers_et_al:2024}. These structural issues obscure the intrinsic electronic signatures of NRSS, making it difficult to disentangle the material's intrinsic properties from extrinsic effects. For instance, Spin- and Angle-Resolved Photoemission Spectroscopy (spin-ARPES), ostensibly the most direct tool for observing NRSS, provides access to spin-split constant energy surfaces across the BZ. However, it demands homogeneity over beam spot sizes of 50~$\mu$m and high-quality stoichiometric samples~\cite{Krempasky_et_al:2024, Reimers_et_al:2024}. Moreover, while spin-ARPES can confirm spin splitting, distinguishing NRSS from conventional spin-orbit coupling effects often requires corroboration through theoretical models. For example, spin-split band features measured in MnTe~\cite{Krempasky_et_al:2024} and $\rm{KV_2Se_2O}$\cite{Jiang2025_inverselieb} were validated by symmetry analysis and tight-binding and/or density functional theory (DFT) calculations to distinguish NRSS from coexisting mechanisms such as relativistic spin splitting (RSS) \cite{bawden_spinvalley_2016, Cao_out_of_plane_rashba_2025}  or ferromagnetism~\cite{Hariki2024}.

In addition to challenges with sample quality and the inherent limitations of spin-ARPES, many experimental techniques yield only indirect evidence of NRSS, complicating efforts to unambiguously attribute observed spin splitting to its true origins. Transport and optical measurements, such as anomalous Hall effect (AHE)~\cite{Feng2022, Gonzalez2023} and circular dichroism~\cite{Hariki2024}, are sensitive to NRSS but often require coexisting ferromagnetism or applied magnetic fields to detect spin splitting. The AHE, for example, typically arises from nonzero magnetization~\cite{Hariki_xmcd_ruo2}, conflicting with the strictly AFM nature of NRSS materials. Similarly, optical probes measure off-diagonal conductivity tensor components that can result from other time-reversal symmetric mechanisms~\cite{sunko_linear_2023}, making their interpretation less definitive. Furthermore, transport signatures depend on splitting at the Fermi level, yet many NRSS metals exhibit their strongest signatures at higher energies~\cite{Yuan_et_al:2020}, limiting the general utility of these techniques. Low N\'eel temperatures in several altermagnetic candidates present another obstacle~\cite{Smejkal2022} since many state-of-the-art measurement tools combining spin-sensitivity with spectromicroscopy can be challenging at low temperatures~\cite{Vasilyev2016_speels, Reyes2023_magnon_tem}. This restricts the conditions under which NRSS can be experimentally verified and complicates efforts to link observed phenomena to theoretical predictions. A fully spin-resolved mapping of both occupied and unoccupied bands is therefore critical for definitively establishing NRSS beyond these indirect methods, underscoring the impact of approaches that measure the entire electronic structure within a single experimental framework.

In this work, we address these challenges by employing a multifaceted approach to directly predict, observe, and understand RSS and NRSS in a \emph{g}-wave altermagnetic candidate, CoNb$_4$Se$_8$, an intercalated transition metal dichalcogenide (TMD). Through symmetry analysis, density functional theory (DFT), and a minimal tight-binding model, we demonstrate precisely how the coupling of crystal field and Zeeman splittings drives NRSS, while the coupling of spin orbit coupling (SOC) and inversion symmetry breaking drives RSS. These models provide the key symmetry conditions for observing RSS and NRSS in this system throughout its electronic structure. We then use these theoretical predictions as a guide for characterization, combining spin-ARPES with a newly developed spin- and angle-resolved electron reflection spectroscopy (spin-ARRES)~\cite{Jobst2015}, thereby probing both occupied and unoccupied energy levels for the first time in a single system.  Importantly, we distinguish observed RSS from NRSS via distinct symmetry-enforced predictions of our model and distinct temperature dependencies across the Néel transition.

\section*{Results}
\begin{figure}
\includegraphics[width=1.\textwidth]{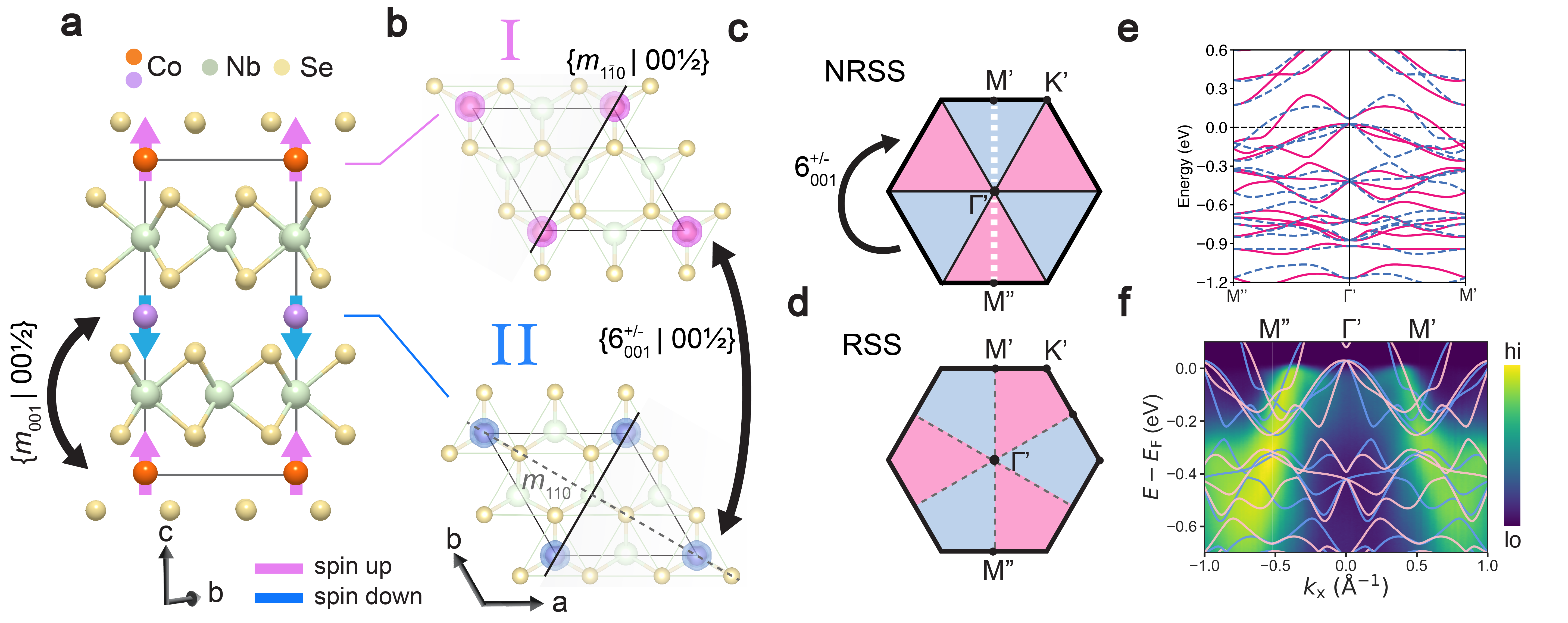}
\caption{\textbf{Real-space magnetic structure of CoNb$_4$Se$_8$}. \textbf{(A)} Magnetic structure of $\mathrm{CoNb_4Se_8}$ along the $a$-axis, and \textbf{(B)} along the $c$-axis at $c$ = 0 (sublattice I) and $c$ = 1/2 (sublattice II). The DFT-calculated pink and blue magnetization clouds correspond to spin up and down, respectively, indicating the $\{6^{\pm}_{001}\,\vert\, 0\, 0 \, \tfrac{1}{2}\}$ screw relating the two sublattices. We show only the bottom Se atoms for clarity.  Solid black (dashed grey) lines correspond to glide-reflection (mirror) planes connecting opposite (identical) spin sublattices. \textbf{(C,D)} Momentum-space spin splitting behavior for NRSS (\textbf{B-}) and RSS (\textbf{C-}) predicted by symmetry analysis, for $k_z \ne 0$. Nodal planes, across which the sign of NRSS (RSS) flips, and corresponding mirror planes in \textbf{B}, are indicated by the solid black (dashed grey) lines. \textbf{(E,F)} Electronic Structure from DFT (\textbf{E-}) and measured in ARPES (\textbf{F-}) along the direction of dotted white line in (\textbf{C}).}
\label{fig:crystalstructure}
\end{figure}

\subsection*{Crystal Structure and Magnetic Order}

$\mathrm{CoNb_4Se_8}$ crystallizes in the hexagonal space group $P6_3/mmc$ (No.~194), with two formula units per unit cell, as shown in Fig.~1A,~1B. In this 1/4-intercalated TMD, $\mathrm{Co}$ ions are positioned between two 2H-$\mathrm{NbSe_2}$ monolayers in octahedral sites, resulting in a centrosymmetric system with collinear AFM order. Unlike the non-centrosymmetric 1/3-intercalated TMD $\mathrm{CoNb_3Se_6}$, which exhibits significant Dzyaloshinskii-Moriya interaction with spins canting away from the $c$-axis~\cite{Xie2024_TMDs, cheong_altermagnetism_2024}, $\mathrm{CoNb_4Se_8}$ maintains robust collinear AFM order~\cite{regmi2024altermag_arxiv}. Although several factors can influence the intercalant configuration in TMDs~\cite{Craig_et_al:2024}, well-ordered crystals of $\mathrm{CoNb_4Se_8}$ have recently been synthesized~\cite{regmi2024altermag_arxiv}, enabling the high-quality samples necessary for our spin-resolved spectroscopies.

DFT calculations confirm that the AFM ground state of $\mathrm{CoNb_4Se_8}$ is energetically favored, lying 60~meV per formula unit lower than the FM state, in agreement with Ref. \cite{regmi2024altermag_arxiv}. The calculated local magnetic moment of $1.49 \, \mu_B$ per Co ion is consistent with a nominal Co-$d^8$ configuration ($S=1$, $m \sim 2\mu_B$) for an itinerant system~\cite{regmi2024altermag_arxiv}. 


The two magnetic sublattices in $\mathrm{CoNb_4Se_8}$ are connected by 12 symmetry operations, including $\{6^{\pm}_{001}\,\vert\, 0\, 0 \, \tfrac{1}{2}\}$ (a 6-fold screw along the crystallographic $c$-axis), $\{m_{001} \,\vert\, 0\, 0 \, \frac{1}{2}\}$ (a reflection across the $ab$-plane followed by a translation along $c$), and a triplet of glides along $c$---$\{m_{1\bar{1}0} \,\vert\, 0\, 0 \, \frac{1}{2}\}$, $\{m_{210} \,\vert\, 0\, 0 \, \frac{1}{2}\}$, and $\{m_{120} \,\vert\, 0\, 0 \, \frac{1}{2}\}$---whose normals are 120$^\circ$ apart. We take the cartesian $z$-axis parallel to the crystal’s $c$-axis by convention (see Fig.~1A,~1B). The two cobalt sites are surrounded by staggered chalcogen environments which can be related to each other by a 6-fold rotation, leading to staggered local crystal field environments and spin densities, as demonstrated by our DFT calculations in (Fig.~\ref{fig:crystalstructure}).

\subsection*{Symmetry and Spin-Splitting in the Electronic Structure}

In the absence of spin-orbit coupling (SOC), the combination of AFM order in $\mathrm{CoNb_4Se_8}$ with its crystalline $6/mmm$ point group results in the spin point group $^26/^2m^2m^1m$, consistent with ``bulk''-type $g$-wave spin-splitting akin to CrSb and MnTe \cite{smejkal_beyond_2022}. Thus, in the limit of zero SOC, we expect spin-split bands everywhere in reciprocal space except at a finite number of nodal planes given by the sublattice-transposing mirrors: the eight planes forming the hexagonal BZ boundary and four additional planes passing through the $\Gamma$-point, namely the $k_z = 0$ plane (enforced by $\{m_{001}\,\vert\,0 0 \frac{1}{2}\}$, which we refer to as $m_{001}$ when discussing $k$-space since the translation is irrelevant), and the three planes containing the $\Gamma - \rm{K}$ high-symmetry lines and orthogonal to the $k_z = 0$ plane (enforced by the triplet of glides-reflections $m_{1\bar{1}0}$, $m_{210}$, and $m_{120}$). Away from these nonrelativistic nodal planes, we generically expect spin-split bands above and below the Fermi level, with larger splitting in bands with a dominant Co orbital character. The application of any of the 12 symmetry operations connecting the opposite-spin sublattices in reciprocal space  (e.g., $6^\pm_{001}$ and $m_{001}$) would lead to a spin flip in the bands.

However, spin-orbit coupling cannot be entirely neglected in $\mathrm{CoNb_4Se_8}$ due to the heavier Nb and Se ions. The inclusion of SOC gives rise to relativistic effects that coexist with NRSS and leads to complex spin textures. 
One such relativistic effect is Rashba-type spin-splitting, which generally develops in  response to surface electric fields produced by broken inversion symmetry  \cite{cao_2025_inplanerashba}. Another such relativistic effect is the spin-polarization ``hidden'' by local inversion symmetry between the two NbSe$_2$ layers, endowing the degenerate bands with an alternating out-of-plane spin texture \cite{zhang_hidden_2014}. This degeneracy can be lifted and RSS developed if the inversion symmetry between the two NbSe$_2$ layers is broken, as occurring naturally at the surface termination. Surface-sensitive probes, including those employed in this work, are more sensitive to one NbSe$_2$ layer than the other and could therefore detect such hidden spin-polarization, 
particularly in bands with large Nb and Se character. This leads to the common spin-valley locked splitting --with alternating spin textures at the $\rm{K}$-points-- as experimentally observed in ARPES measurements of the nonmagnetic parent compound NbSe$_2$ \cite{bawden_spinvalley_2016}. 
The key aspect to experimentally distinguish nonrelativistic from relativistic splittings is that the RSS originating from the NbSe$_2$ structure presents spin degenerate nodal planes containing the $\Gamma - \rm{M}$ paths (due to the mirror symmetries of the crystal structure , Fig.~\ref{fig:crystalstructure}B, ~\ref{fig:crystalstructure}D) while allowing splitting at the $\rm{K}$ points. This suggests that any spin splitting near the $\rm{K}$ ($\rm{M}$) points is of relativistic (nonrelativistic) origin.

Here, we employ a combination of symmetry analysis and theoretical calculations, including DFT and intuitive minimal models, to disentangle the experimental signatures of NRSS and RSS.


\subsection*{Crystal field-driven NRSS}

\begin{figure}[t]
\includegraphics[width=1.\textwidth]{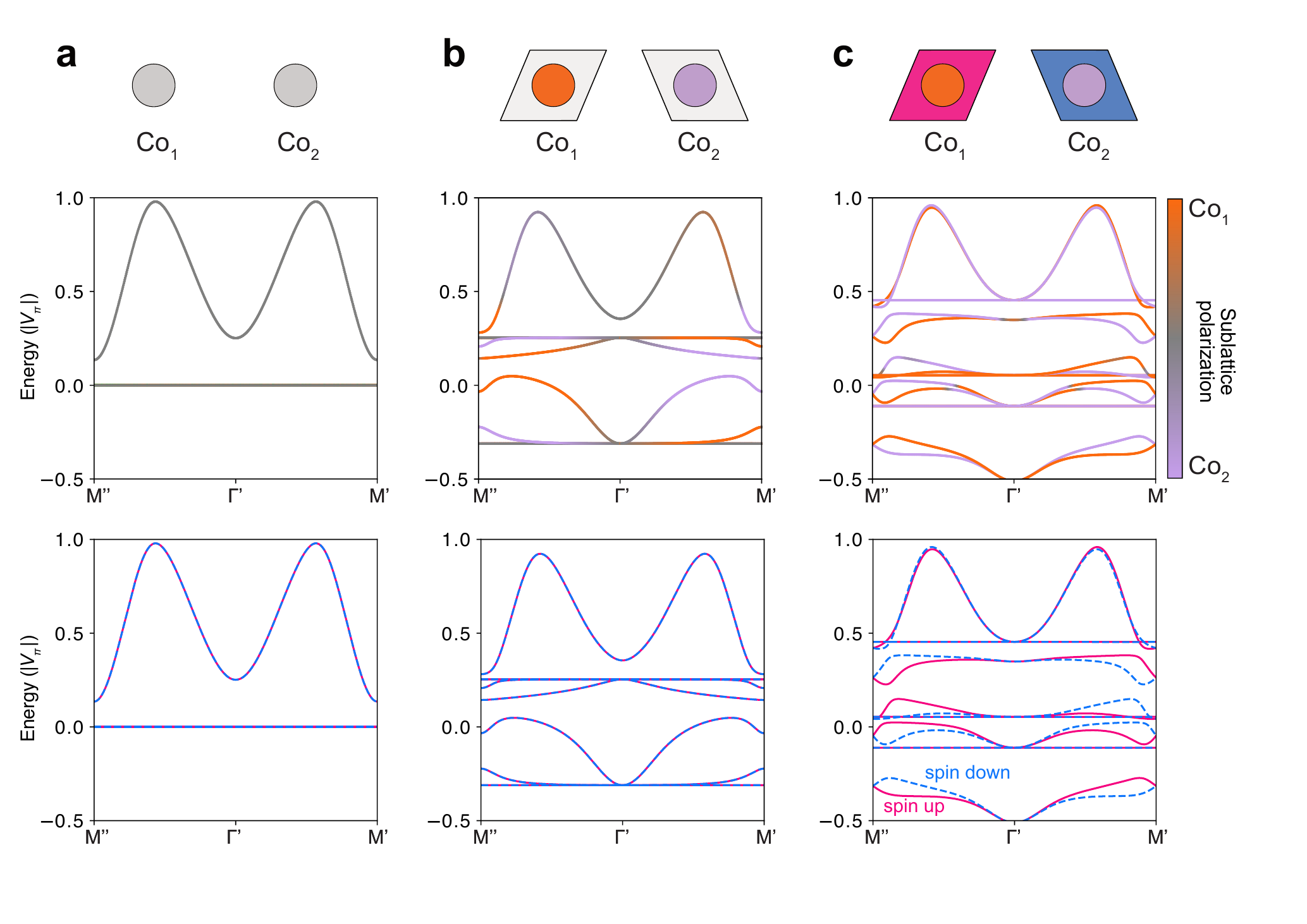}
\caption{\textbf{Tight-binding band structure of NRSS} \textbf{(A)} Results of the model without including CF and exchange terms, indicated with bare Co sites. Top (bottom) row of the bands shows the sublattice polarization (spin splitting). In both cases, the bands are unpolarized. \textbf{(B)} Results where CF is included, indicated with two octahedrons related by the screw operation $\{6^\pm_{001} | 0\,0\,  \frac{1}{2}\}$. Sublattice polarization arises, while spin splitting remains zero. \textbf{(C)} Results where both CF and exchange interactions are included, indicated by Co sites colored with pink and blue denoting up and down magnetization, respectively. Here, sublattice polarization remains finite and NRSS also develops.}
\label{fig:TB}
\end{figure}

The microscopic mechanism behind NRSS can be captured by a minimal tight-binding model that includes both \textit{d} orbitals of the magnetic sublattices and \textit{p} orbitals of the ligand sites. It has the form
\[
\hat{\mathcal{H}} = \hat{\mathcal{H}}_0 + \hat{\mathcal{H}}_{CF} + \hat{\mathcal{H}}_{ex},
\]
where $\hat{\mathcal{H}}_0$ is the hopping between magnetic and nonmagnetic sites, $\hat{\mathcal{H}}_{CF}$ captures the explicit effects of the $D_{3d}$ crystal field of each magnetic sublattice, and $\hat{\mathcal{H}}_{ex}$ introduces an effective collinear, antiferroic magnetic order. The explicit definition of each term is detailed in the SM, and here we reveal how the interplay of these terms gives rise to NRSS. When only $\hat{\mathcal{H}}_0$ is included, the hopping interaction between magnetic ions and ligands alone does not produce NRSS, as seen in Fig. \ref{fig:TB}A (bottom row). Moreover, plotting the sublattice projection (Fig. \ref{fig:TB}A, middle row) shows that electrons are also sublattice unpolarized.

When the crystal field is turned on (Fig. \ref{fig:TB}B), although spin splitting is still absent, bands that were degenerate now split into sublattice-polarized bands. This occurs because Co$_1$ and Co$_2$ sites have different crystal environments along a specific real-space direction, which translates into reciprocal-space paths. Importantly, such polarization reverses between $6^\pm_{001}$-related momentum directions ($\Gamma - \mathrm{M^{\prime\prime}}$  
and $\Gamma - \mathrm{M^{\prime}}$) and such a feature is a key precondition for NRSS.

Finally, when the antiferroic exchange interaction is included, each sublattice experience a magnetization of opposite sign. Because bands related by $6^\pm_{001}$ symmetry have precisely opposite sublattice polarization, such bands develop an opposite Zeeman splitting, thus leading to the NRSS seen in Fig. \ref{fig:TB}C.

Our model and symmetry analysis, guided and supported by DFT results, establish a comprehensive theoretical framework for understanding NRSS in $\mathrm{CoNb_4Se_8}$. Using our DFT and theoretical results predicting those regions of the BZ with NRSS, we now present clear evidence of RSS and NRSS in the spin-resolved electronic structure above and below the Fermi level ($E_{\rm{F}}$).

\begin{figure}
\includegraphics[width=1.\textwidth]{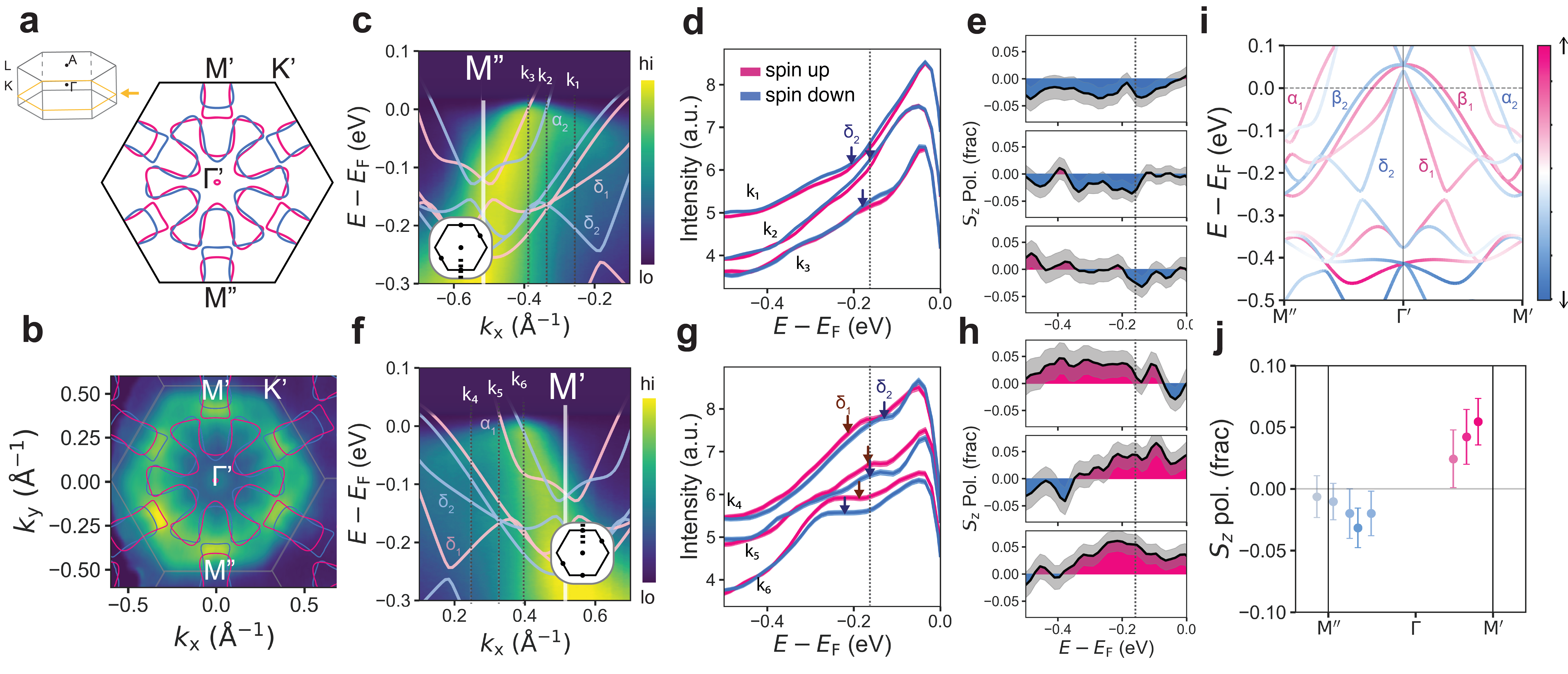}
\caption{\textbf{ARPES measurement of nonrelativistic spin-splitting in occupied electronic structure in CoNb$_4$Se$_8$.} \textbf{(A,B)} Fermi surface calculated from DFT (\textbf{A-}) and measured in ARPES (\textbf{B-}) for $k_z $ near $0.25c^*$ as indicated by the inset. \textbf{(C)} 13~K ARPES spectra along $\mathrm{M^{\prime\prime} - \Gamma{^\prime}}$ , as indicated by the sketch in the inset. Spin up (down) DFT bands from \textbf{A} are overlaid in pink (blue). \textbf{(D,E)} Spin-resolved EDCs spectra (\textbf{D-}) and corresponding spin polarization along the $z$ axis (\textbf{E-})  at momenta indicated by black vertical lines in \textbf{B}.  Measurement uncertainty out to to $1\sigma$ outlined in grey. Red (navy) arrows indicate DFT energies of the $\delta_1$($\delta_2$) band.  \textbf{(F-H)} Same as (\textbf{C-E}) but for electronic structure along  $\mathrm{ \Gamma{^\prime} -M^{\prime}} \equiv (6^{\pm}_{001})^3(\mathrm{M^{\prime\prime} - \Gamma{^\prime}})$ direction in momentum space. \textbf{(I)} Spin-resolved nonrelativistic DFT calculations of electronic structure along  $\mathrm{M^{\prime\prime} - \Gamma{^\prime} - M^{\prime}}$ projected on the Co ions. The color indicates cobalt contribution to each band multiplied by $\pm1$ for spin up and down. \textbf{(J)} $S_\mathrm{z}$ polarization as a function of momentum along $\mathrm{M^{\prime\prime} - \Gamma{^\prime} - M^{\prime}}$, extracted from binding energy depicted by dotted lines in (\textbf{E,H}).
}
\label{fig:arpes_mgm}
\end{figure}

\subsection*{NRSS in Occupied Electronic Structure of CoNb$_4$Se$_8$ via ARPES}

First principles calculations and angle-resolved photoemission spectroscopy (ARPES) measurements of $\mathrm{CoNb_4Se_8}$ (Fig.\ref{fig:arpes_mgm}) capture the NRSS enforced by crystal symmetry. 

Away from the spin group symmetry enforced nodal plane at $k_z = 
0$ , the nonrelativistic DFT-calculated Fermi surface of $\mathrm{CoNb_4Se_8}$ (Fig. \ref{fig:arpes_mgm}A) contains pairs of threefold-symmetric spin polarized Fermi pockets:  notably petal-shaped ones surrounding the zone center ($\Gamma'$) and lobe-shaped ones surrounding the zone edge ($\mathrm{M^{\prime}}$). As is clear by eye, spin up pockets at parallel momentum $\vec{k_\parallel}$ map to spin down pockets located at  $6^{\pm}_{001}(\vec{k_\parallel})$, as well as those at $-\vec{k_\parallel} = (6^{\pm}_{001})^3(\vec{k_\parallel})$, consistent with NRSS behavior. The raw ARPES Fermi surface spectra in (Fig. \ref{fig:arpes_mgm}B)  also presents pairs of threefold symmetric Fermi pockets which are petal shaped at zone center and lobe-shaped at zone edges, consistent with the overlaid DFT calculations and with previous reports on this material\cite{devita2025opticalswitchinglayeredaltermagnet,sakhya2025electronicstructurelayeredaltermagnetic}. 

Along the momentum path connecting zone edges ($\mathrm{M^{\prime\prime} - \Gamma{^\prime}}$ and $\mathrm{\Gamma{^\prime} - M^{\prime}}$ directions in momentum space),  the DFT-calculated spin-polarized electronic band structure (Fig.\ref{fig:arpes_mgm}I) reveals explicit spin splitting throughout the occupied and unoccupied states. Prominent band features include steep electron pockets originating from zone edge at $\approx E_
\mathrm{F}-100\,\mathrm{meV}$, labeled $\alpha_{1,2}$,  hole-like bands extending from just above $E_
\mathrm{F}$ at zone center to $\approx E_
\mathrm{F}-250\,\mathrm{meV}$ at zone edge, labelled $\beta_{1,2}$, and steep hole bands at zone center that flatten near  $\approx E_
\mathrm{F}-200\,\mathrm{meV}$ at zone edge, labeled  $\delta_{1,2}$ . As is clear by eye, the pair of  ${\delta}$ bands host 100 meV-scale spin splitting that flips upon reversing parallel momentum, consistent with NRSS behavior.  Raw ARPES data along the $\mathrm{M^{\prime\prime} - \Gamma{^\prime}}$ and $\mathrm{\Gamma{^\prime} - M^{\prime}}$ directions in momentum space (Fig.\ref{fig:arpes_mgm}C,F) confirm the existence of these features, with matrix elements favoring the $\alpha$ pockets at the zone edges over the fainter but still visible $\delta$ bands. 

Importantly, spin-resolved energy distribution curves (EDCs) spectra along $\mathrm{M^{\prime\prime} - \Gamma{^\prime}}$ (Fig.\ref{fig:arpes_mgm}D) confirm features  $\approx$170 meV below $E_\mathrm{F}$, which have a faint but negative spin polarization (Fig.\ref{fig:arpes_mgm}E) along the $z$ direction. These spin polarized features are in qualitative agreement with the positions of the $\delta_2$ bands predicted by DFT. The corresponding features along $\mathrm{\Gamma{^\prime} - M^{\prime}}$ (Fig.\ref{fig:arpes_mgm}G,H) have a positive spin polarization at the same energy. The stronger spin up features here correspond qualitatively to the locations of the DFT  $\delta_1$ bands, while the weaker spin down features correspond to the DFT $\delta_2$ bands (See Supplementary Material). The momentum dependence of the spin polarization, summarized in Fig. \ref{fig:arpes_mgm}J, therefore flips sign across the zone center. While the DFT calculation presents two spin split $\delta$ bands, the experiment resolves one spin polarized band ( $\downarrow$ along $\mathrm{M^{\prime\prime} - \Gamma{^\prime}}$ and $\uparrow$ along $\mathrm{\Gamma{^\prime} - M^{\prime}}$) more strongly than its spin-split counterpart. Such matrix elements effects can occur when the photoemission experiment is more sensitive to states from a single Co sublattice (see Supplemental Material). Such alternating spin polarization, measured here at 13K, is only present below the N\'eel ordering temperature. Above the N\'eel temperature, this polarization is strongly suppressed, providing strong evidence for NRSS behavior (see Supplemental Material).  
 
Altogether, these results identify $\mathrm{CoNb_4Se_8}$ as a bulk-type \textit{g}-wave altermagnet, consistent with other reports (\cite{devita2025opticalswitchinglayeredaltermagnet,sakhya2025electronicstructurelayeredaltermagnetic}) and indicate similar properties to \textit{g}-wave compounds MnTe~\cite{Krempasky_et_al:2024} and CrSb~\cite{Reimers_et_al:2024}, which share the same spin point group. 

\begin{figure}
\includegraphics[width=1.\textwidth]{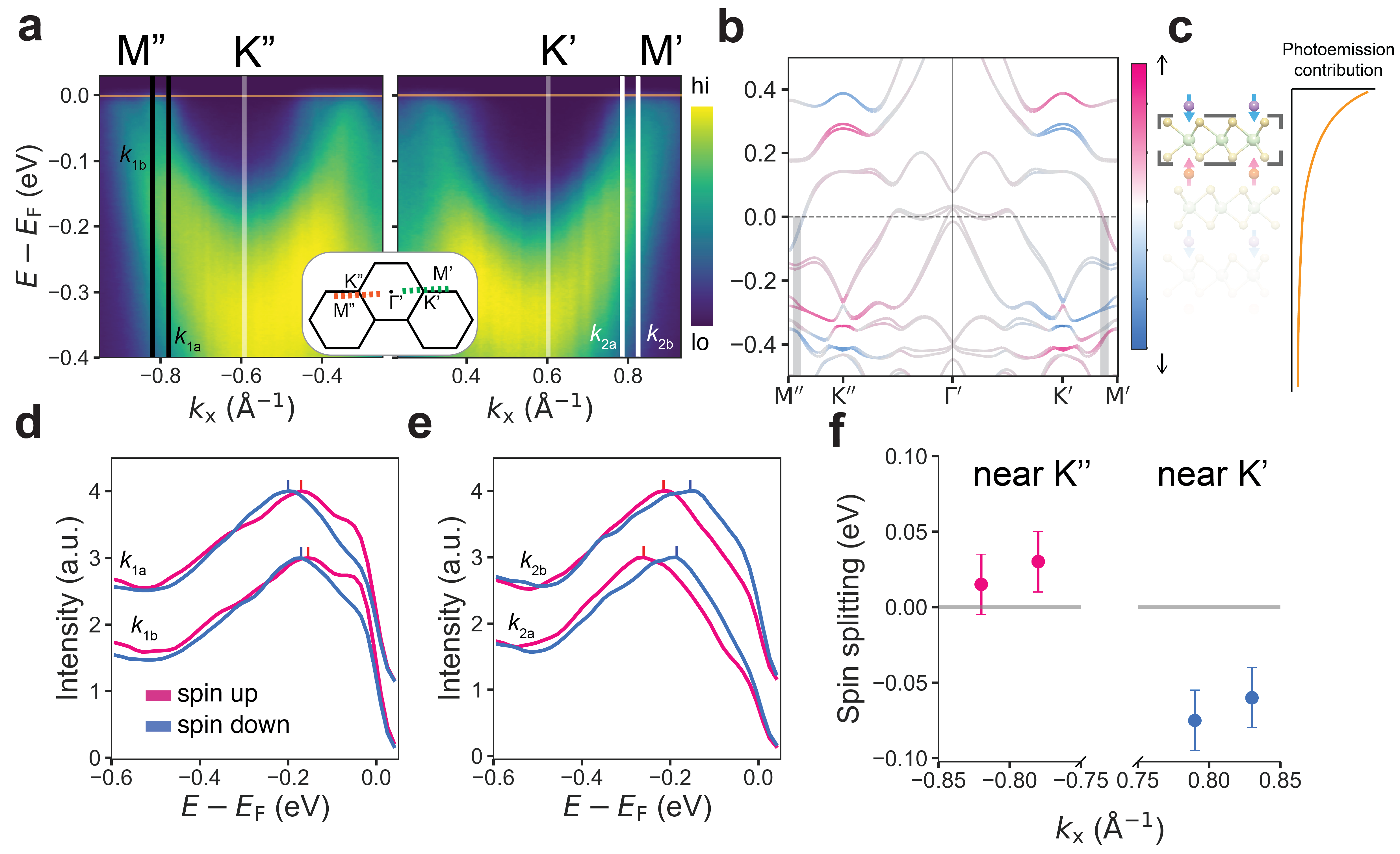}
\caption{\textbf{ARPES measurement of relativistic spin-splitting in occupied electronic structure in CoNb$_4$Se$_8$.} \textbf{(A)} 13~K ARPES spectra along electron pockets surrounding $\mathrm{K^{\prime\prime}} \equiv (6^{\pm}_{001})^3 \mathrm{K^{\prime}}$ (left) and $\mathrm{K^{\prime}}$ (right) as indicated by the inset momentum space cartoon. (\textbf{B}) First-principles calculation of spin-resolved band structure projected on one $\mathrm{NbSe_2}$ layer (\textbf{C}) Sketch of photoelectron probe depth  governed by electron mean-free path at $h\nu = 55\, \rm{eV}$. Grey dashed rectangle indicates one NbSe$_2$ layer.  (\textbf{D, E}) Spin-resolved EDCs spectra along momenta indicated by black (white) vertical lines in \textbf{A}. Red (blue) ticks indicate peak locations for bands polarized spin up (down), extracted as a parabolic fit within the vicinity of spectral maximum. (\textbf{F}) Representative momentum-dependent spin splitting, extracted from the difference in spin up and spin down band locations in \textbf{D} and \textbf{E} . 
}
\label{fig:arpes_kgk}
\end{figure}

\subsection*{RSS in Occupied Electronic Structure of CoNb$_4$Se$_8$ via ARPES}


Figure  \ref{fig:arpes_kgk} presents spin-resolved first-principles calculations and photoemission spectra along $\rm{K^{\prime\prime} - \Gamma^{\prime} - K^{\prime}}$, parallel to a nodal plane enforced by nonrelativistic spin group symmetries. Interestingly, the relativistic DFT band structure projected on one NbSe$_2$ layer in Fig. \ref{fig:arpes_kgk}B  presents finite spin splitting in $\mathrm{K^{\prime\prime}}$  and $\mathrm{K^{\prime}}$ valleys. Such splitting is finite only when spin-orbit coupling is present, and is equal and opposite when the bands are projected on the other NbSe$_2$ layer (see Supplemental Material). This behavior is consistent with a relativistic spin polarization driven by local inversion symmetry breaking described earlier in the text \cite{zhang_hidden_2014,bawden_spinvalley_2016}, and cancels out when both layers are accounted for.

Experimentally observing said RSS requires spin- and sublattice- resolved electronic structure measurements. ARPES probes materials at the surface termination, where inversion symmetry is broken. By choosing a photon energy of $h\nu$ = 55 eV, only the top NbSe$_2$ layer provides the dominant contribution to the photoemission experiment, as indicated by the sketch in Fig. \ref{fig:arpes_kgk}C\cite{Hufner:2013}. 

Raw ARPES data along the $\mathrm{M^{\prime\prime} - K^{\prime\prime} - \Gamma^{\prime}}$ and $\mathrm{\Gamma{^\prime} - K^{\prime}-M^{\prime}}$ directions in momentum space (Fig.\ref{fig:arpes_kgk}A) reveal two electron pockets near the BZ corner with a depth of $\approx 0.35 \,~\mathrm{eV}$, similar to those predicted by DFT.  Spin-resolved EDCs spectra around the $\rm{K^{\prime\prime}}$ pocket (Fig.~\ref{fig:arpes_kgk}D) display significant spin polarization. Focusing on the strongest spectral peaks $\approx$200 meV below the Fermi level, we see that states polarized along the +$c$ direction are shifted closer to the Fermi level by 30 $\pm$ 20~meV than those along $-c$. By contrast, in the vicinity of the $\rm{K^{\prime}}$ pocket at momenta $k_{2a} \approx 2^{\pm}_{001}(k_{1a})$ (Fig.~\ref{fig:arpes_kgk}E), the spin splitting is inverted: states polarized along +c are now 70 $\pm$ 20 meV further from $E_{\rm{F}}$ . Fig.~\ref{fig:arpes_kgk}F summarizes this alternating spin splitting, which is qualitatively consistent with the RSS predicted by the DFT calculations, and quantitatively similar to relativistic spin-momentum locking observed in the topmost layer of parent material $\rm{NbSe_2}$\cite{bawden_spinvalley_2016}.

\begin{figure}
\includegraphics[width=1.\textwidth]{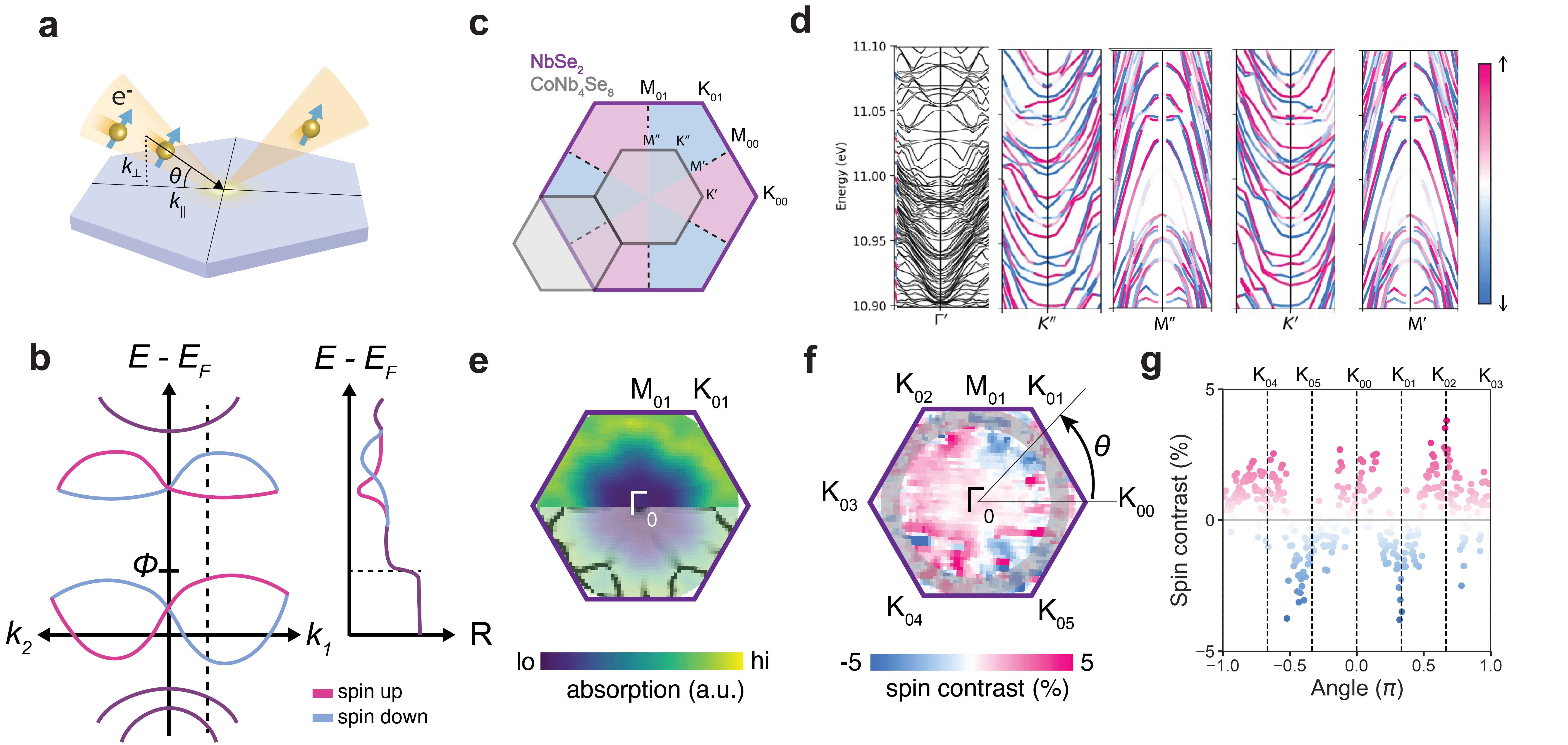} 
 \caption{\textbf{ARRES measurement of spin-split unoccupied electronic structure in CoNb$_4$Se$_8$} \textbf{(A, B)}  Schematic of sp-ARRES experiment: Spin-polarized electrons incident on the sample are preferentially absorbed if there is an unoccupied electronic state with corresponding spin $\vec{s}$, energy $E$ and momentum $\vec{k}$.  \textbf{(C)} Schematic of BZ unfolding in CoNb$_4$Se$_8$ onto the parent BZ of NbSe$_2$.
 \textbf{(D) }DFT calculations of spin-resolved unoccupied electronic structure near high symmetry points 11 eV above $E_\mathrm{F}$.  \textbf{(E, F)} Spin-integrated \textbf{(E)} and spin-polarized \textbf{(F)} ARRES spectra at constant energy surface 11~eV above $E_\mathrm{F}$ at T = 30~K. Band locations, indicated by maximal band curvature, overlaid in black. \textbf{(G)} Spin contrast in the ARRES experiment along the region indicated by the grey ring in \textbf{F} centered at $|k_\parallel|= 0.8|\rm{\bar{\Gamma M}}|_{\mathrm{NbSe_2}}$.  
}
\label{fig:spleem}
\end{figure}

\subsection*{The Unoccupied Electronic Structure of CoNb$_4$Se$_8$ via ARRES}

RSS and NRSS in $\mathrm{CoNb_4Se_8}$ are not restricted to the quasiparticle states near $E_{\rm{F}}$ but extend over a large energy range. In Fig. \ref{fig:spleem}, we compare calculated and measured spin- resolved electronic structure 11~eV above $E_{\rm{F}}$. 

We introduce a new technique called spin- and angle-resolved reflected electron spectroscopy (spin-ARRES) (Fig.~\ref{fig:spleem}A), which improves on methods for probing the unoccupied electronic structure~\cite{strocov_very_2000,Jobst2015} by adding the spatial and spin resolution available to SPLEEM~\cite{rougemaille_magnetic_2010} . Spin-ARRES measures the reflectivity of the sample surface for an incident electron with energy \( E \), momentum \( \vec{k} \), and spin \( \vec{s} \) (Fig.~\ref{fig:spleem}B). The reflectivity sharply drops when an unoccupied electronic state is present at \( (E, \vec{k}, \vec{s}) \) in its electronic structure (see Methods for more details). Analogous to spin- and angle-resolved inverse photoemission spectroscopy (SP-ARIPES)~\cite{donath_spin-resolved_1989, Campos2022}, spin-ARRES offers a significantly greater interaction cross-section (by a factor of \( 10^5 \)) due to its use of electrons rather than photons as a probe, enabling higher-throughput spectroscopy~\cite{Woodruff:2016}.

The spin-integrated ARRES constant energy surface for $\mathrm{CoNb_4Se_8}$, shown in Fig.~\ref{fig:spleem}E, reveals a clear band-like structure in the unoccupied energies well above \( E_F \). The electronic absorption curvature~\cite{Zhang_2011_arpescurvature}, outlined in black, peaks at momentum regions corresponding to unoccupied states. These features exhibit the periodicity of the parent $\rm{NbSe_2}$ structure, often observed in bands lacking intercalant orbital character (See Supplementary Material)\cite{popcevic_conb3s6_arpes}.

First-principles calculations with a slab geometry, presented in Fig.  \ref{fig:spleem}D, indicate that at energies 11eV above the Fermi level, there exists meV-scale spin-splitting with \( \vec{s}\parallel c \) that alternates between $\mathrm{K^{\prime\prime}}$ and $\mathrm{K^{\prime}}$ valleys of the CoNb$_4$Se$_8$ BZ, and is suppressed at the $\mathrm{M'}$ and $\mathrm{M''}$ points.  In contrast to the layer-projected spin-momentum locking present in the occupied states, here the spin splitting is only present in a slab geometry with finite spin orbit coupling.  Following symmetry arguments from the previous sections, such spin splitting is of relativistic origin, likely from a Rashba-effect which forms from the in-plane electric fields present in individual layers of 2H transition metal dichalcogenides away from the mirror planes equivalent to $m_{110}$  \cite{cao_2025_inplanerashba}. 

The surface sensitivity of ARRES enables observation of energy splitting in surface bands\cite{Jobst2015}. Just as electron reflectance contrast \( \frac{I_\uparrow - I_\downarrow}{P (I_\uparrow + I_\downarrow)} \) indicates spin-splitting of unoccupied electronic states at \( \vec{k_\parallel} = 0 \) in SPLEEM~\cite{rougemaille_magnetic_2010}, here in spin-ARRES the contrast indicates energy splitting of electronic states at \( \vec{k_\parallel} \neq 0 \). The spin-ARRES spectra with \( \vec{s}\parallel c \) in Fig.~\ref{fig:spleem}F present a spin texture that has suppressed contrast at the \( \Gamma^\prime \) point, with finite spin contrast appearing predominantly at \( \vec{k} \neq 0 \) that alternates with sixfold symmetry.  Indeed, the integrated spin contrast at a radius of $0.8|\rm{\bar{\Gamma M}}|_{\mathrm{NbSe_2}}$, presented in Fig.~\ref{fig:spleem}G, is largest around the $\mathrm{K}_{0i}$ pockets and smallest near $\mathrm{M}_{0i}$ pockets of the NbSe$_2$ parent BZ. As NbSe$_2$ hosts the same mirror symmetries as $\mathrm{CoNb_4Se_8}$, the observed sixfold symmetric spin contrast is consistent with the predictions from the slab DFT calculations (see Fig. \ref{fig:spleem}D), indicating the presence of RSS in the unoccupied electronic band structure of $\mathrm{CoNb_4Se_8}$. 


 
\section*{Discussion} 
This study represents a pivotal step in establishing CoNb$_4$Se$_8$ as a prototypical \emph{g}-wave altermagnet, offering a robust framework for understanding NRSS and distinguishing it from RSS. By combining theory and calculations with complementary experimental probes -- spin-ARPES and spin-ARRES -- we directly observe symmetry-enforced NRSS and inversion symmetry-breaking-induced RSS. These findings address longstanding challenges in observing NRSS and uncover new physics tied to symmetry-protected spin-split states in quantum materials.

The complementary use of spin-ARPES and spin-ARRES to measure the electronic structure of CoNb$_4$Se$_8$ offers unprecedented insight into the nature of NRSS in CoNb$_4$Se$_8$, and addresses key limitations of existing probes. Spin-ARPES directly corroborates the predicted spin-split bands in the occupied electronic structure, providing momentum-resolved evidence of \emph{g}-wave symmetry, and validating the predictions of our theoretical framework. Meanwhile, spin-ARRES extends this capability to unoccupied states, revealing RSS across a broad energy range and overcoming the energy limitations of spin-ARPES. The combination of these techniques overcomes the limitations of single-method approaches, where spin-ARPES alone may struggle with resolution and spatial inhomogeneity, and indirect probes such as AHE measurements are constrained by coexisting ferromagnetism or the need for applied magnetic fields.

Our introduction of spin-ARRES as a novel probe of the unoccupied spin-polarized band structure offers complementary information to spin-ARPES. Despite having a broader energy resolution ($\sim$100~meV; see Supplementary Information) that limits its ability to distinguish NRSS at fine energy scales, its superior spatial resolution ($\sim$1~$\mu$m) over ARPES opens exciting possibilities. For example, the technique enables high-resolution magnetostructural measurements of altermagnetic domains on the meso- and nanoscale, which is crucial for applications requiring localized control over spin states in spintronic devices~\cite{Jobst2015,Amin2024}. Indeed, the relatively small spin polarizations ($\approx5 \%)$ measured here in ARPES could indicate the presence of altermagnetic domains smaller than the ARPES probe size ($\approx100\mu$m)\cite{devita2025opticalswitchinglayeredaltermagnet, Amin2024}. Additionally, our experimental demonstration of RSS of unoccupied bands opens the door to a wealth of optical signatures of NRSS in altermagnets where the splitting of both occupied and unoccupied bands is required~\cite{Vila2024, Adamantopoulos2024, Weber2024}.


Crucially, we observe a clear altermagnetic phase transition in $\mathrm{CoNb_4Se_8}$, with the suppression of NRSS above the Néel temperature, confirming that the alternating spin splitting along $\mathrm{M^{\prime\prime}-\Gamma{^\prime} - M^{\prime}}$  is directly linked to altermagnetic order (See Supplementary Material). This temperature-dependent evolution, combined with the alternating character across the zone center, distinguishes the observed spin splitting from conventional mechanisms such as spin-orbit coupling, and less conventional mirror-even spin textures observed recently\cite{din2025_relativistic_bands_altermagnet}. By contrast, the persistence of alternating spin splitting around the $\mathrm{K'}$ valleys above $T_N$ implies said splitting is of relativistic origin (See Supplemental Material). The presence of RSS in $\mathrm{CoNb_4Se_8}$ offers an intriguing materials system to explore the interplay of NRSS with spin-orbit coupling, out of which topological properties can emerge \cite{Smejkal2022,Antonenko_PRL2025,sah2025altermagnetismkagomeflatband}. Indeed, we observe Fermi surface spectral weight delineating the BZ boundary that is not captured by bulk DFT calculations, which may indicate the presence of topological surface states observed in other NRSS materials\cite{li_topological_2025, lu_signature_2025}.

The introduction of NRSS to the phases available in intercalated transition metal dichalcogenides~\cite{regmi2024altermag_arxiv, Mandujano_et_al:2024} offers an inviting materials family in which to explore the interplay of complementary and competing orders in phase transitions. For example, the parent compound 2H-NbSe$_2$ hosts competing CDW~\cite{Rossnagel2011},  anisotropic superconducting~\cite{Fletcher2007_nbse2, Rahn2012_nbse2}, and pair density wave order~\cite{Liu2021_pdw}.  Though $\mathrm{CoNb_4Se_8}$ presented in this work exhibits no superconducting order, there is some evidence for the 3x3 CDW phase present in the parent compound~\cite{regmi2024altermag_arxiv}. Lifting of time-reversal symmetry in altermagnets therefore applies additional constraints on the pairing symmetry and angular momentum of Cooper pairs  ~\cite{jungwirth_altermagnets_2024}.

The vast playground of possibilities in intercalated TMDs offer a range of possibilities for exploring coexisting and competing phases in altermagnetic TMDs. While this work explores local symmetry breaking in altermagnets, other reports show that NRSS persists upon breaking global inversion symmetry through tuning intercalant concentration\cite{mandujano_evolution_2025} and strain engineering   \cite{din2025_relativistic_bands_altermagnet},  allowing for the coexistence of spin-glass, spin helical textures, and NRSS behavior in the same samples. Altermagnetic TMDs with stronger spin orbit coupling have been synthesized~\cite{Mandujano_et_al:2024}, and the van-der-Waals nature of the material leaves tantalizing possibilities for interface- and heterostructure-enabled quantum phases ~\cite{Cao2018b,heinsdorf_altermagnetic_2024, jungwirth_altermagnets_2024}.

\section*{Acknowledgements}

We acknowledge Arian Gashi for technical assistance in the sample holder fabrication setup, Andreas Schmid for assistance on the QSPLEEM experimental setup. We appreciate helpful discussions about the DFT calculations and crystal field analysis with Ella Banyas, Isaac Craig, Jack Broad, Kevin Moseni, and Veronika Sunko. We thank Igor Mazin for useful discussions, and Aeron Hammack, Shelly Conroy, John Vinson and Adam Schwartzberg for feedback on drafts of this manuscript.

\subsection*{Funding}
This work was primarily supported Laboratory Directed Research and Development Program of LBNL under the U.S. Department of Energy (DoE) Contract No. DE-AC02-05CH11231. The theoretical work was funded by the U.S. Department of Energy, Office of Science, Office of Basic Energy Sciences, Materials Sciences and Engineering Division under Contract No. DE-AC02-05-CH11231 within the Theory of Materials program. Work was performed at the Molecular Foundry and at the Advanced Light Source supported by the Office of Science, Office of Basic Energy Sciences, of the U.S. Department of Energy under contract no. DE-AC02-05CH11231. This research used resources of the National Energy Research Scientific Computing Center, a DOE Office of Science User Facility supported by the Office of Science of the U.S. Department of Energy under Contract No. DE-AC02-05CH11231 using NERSC award BES-ERCAP0020966. N.J.G. and R.B.R. were supported by Army Research Office under Cooperative Agreement Number W911NF- 22-2-0173. 

\subsection*{Author Contributions}
S.M.G., N.D., A.S., and C.W.J. conceived the study. O.A.A., J.F., and N.D. conducted DFT calculations. M.V. developed the tight-binding model with input from O.A.A.,N.D.,J.F., and S.M.G. R.B.R. and N.J.G. grew the crystals. N.D. and A.F. conducted spin-ARPES measurements. C.W.J. developed the ARRES technique at the Foundry, E.S.B. and N.D. developed the software for reproducible spin-ARRES measurements, N.D. and A.S. conducted the spin-ARRES measurements. N.D., O.A.A., M.V., and S.M.G. wrote the paper with input from all coauthors. S.M.G. supervised the project.

\subsection*{Data and Materials Availability}
All data needed to evaluate the conclusions in the paper are present in the paper and/or the Supplementary Materials.

\subsection*{Competing Interests}
The authors declare that they have no competing interests.

\bibliography{nick,marc, Codes, pseudos, Functionals, sinead, spin_group_history}

\section*{Supplementary Material}
\setcounter{figure}{0}
\renewcommand{\thefigure}{S\arabic{figure}}
\renewcommand{\thetable}{S\arabic{table}}
\renewcommand{\theequation}{S\arabic{equation}}

\subsection*{Materials and Experimental Methods}
\subsubsection*{Crystal Growth}

Single crystals of $\mathrm{CoNb_4Se_8}$ were grown~\cite{regmi2024altermag_arxiv} by chemical vapor transport using iodine as the transport agent. First, a polycrystalline sample was prepared by heating stoichiometric amounts of cobalt powder (Alfa Aesar 99.998\%), niobium powder (Alfa Aesar 99.8\%), and selenium pieces (Alfa Aesar 99.9995\%) in an evacuated silica ampule at 950 $^{\circ}$C for 5 days. Subsequently, 2 g of the powder was loaded together with 0.4 g of iodine in a fused silica tube of 14 mm inner diameter. The tube was evacuated and sealed under vacuum. The ampule of 10~cm length was loaded in a horizontal tube furnace in which the temperature of the hot zone was kept at 950 $^{\circ}$C  and that of the cold zone was $\approx$ 850 $^{\circ}$C  for 7 days. Several $\mathrm{CoNb_4Se_8}$ crystals formed with a distinct, well-faceted flat plate-like morphology.

\subsubsection*{spin-ARPES measurements}

ARPES was measured at the Advanced Light Source beamline 10.0.1 using a Scienta R4000 spectrometer equipped with DA30 deflector plates for spin-integrated ARPES mapping and for steering electrons into dual very low energy electron diffraction (VLEED) spin-detectors. Figures in the main text were acquired using a photon energy of 55~eV, a temperature of 13~K, and a pressure better than 5e-11 Torr, producing an overall energy and momentum resolution of 10~meV and 0.01~\AA$^{-1}$, respectively. 

The two exchange-scattering type spin-detectors use in-plane magnetization of FeO thin film targets to provide $\left(k_z, k_x\right)$ and $\left(k_z, k_y\right)$ components of the spin-asymmetry, e.g. with redundancy in the $k_z$ component. For each spin-detector, the sequentially measured spectra $I_{+}(\omega)$ and $I_{-}(\omega)$ are used to compute the raw spin-scattering asymmetry, $A_{ \pm}(\omega)=\left(I_{+}-I_{-}\right) /\left(I_{+}+I_{-}\right)$, which is corrected by the instrumental spin-scattering efficiency factor, i.e. the Sherman function $S_{\text {eff }}$, to determine the photoelectron spin-polarization $P(\omega)=A_{ \pm}(\omega) / S_{e f f}$. The corrected spin-dependent spectra are then calculated as $I_{\uparrow \downarrow}(\omega)=I_{a v}(\omega)(1 \pm P(\omega))$, where $I_{a v}=\left(I_{+}+I_{-}\right) / 2$. The Sherman function for this exchange spin-detector is calibrated to be $S_{\text {eff }} \approx 0.2$.

All ARPES data in this paper were analyzed using pyARPES, an open-source python-based analysis framework~\cite{Stansbury2020}.

\subsubsection*{spin-ARRES measurements}
\begin{figure}
\includegraphics[width=1.\textwidth]{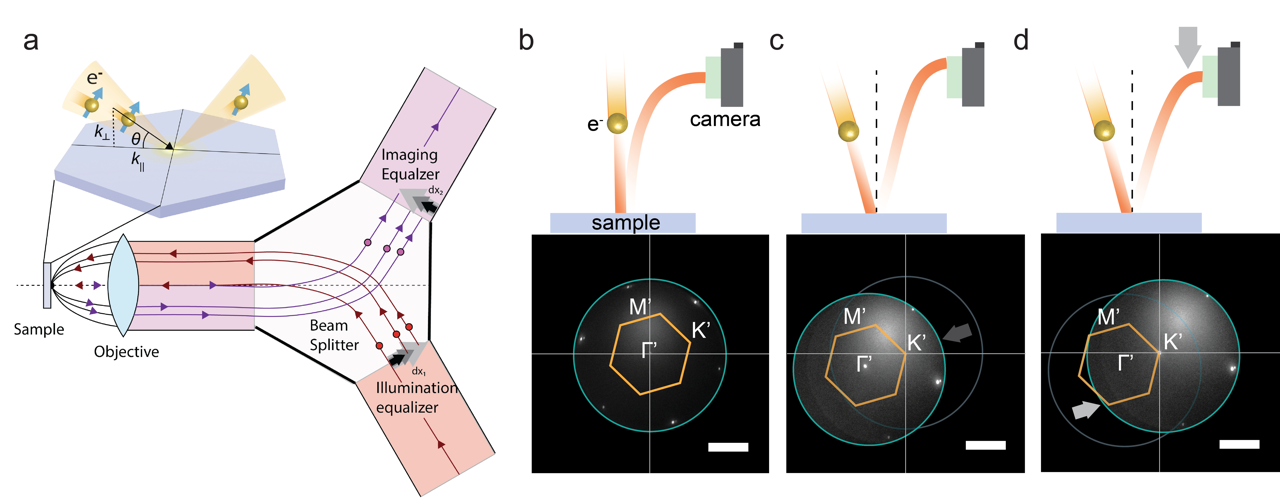}
\caption{\textbf{Angle Resolved Electron Reflection Spectroscopy Technique}. (\textbf{A}) Schematic for tuning parallel momentum of incident electrons in the QSPLEEM. Incident electrons follow the red trajectories through the illumination column into the beam splitter (triangular region) before being focused down by the objective lens onto the sample. Reflected electrons follow the purple trajectories back through the objective lens and beam splitter towards the Imaging column. Trajectories have been simplified,  i.e. they lack image and diffraction planes, for ease of illustration (\textbf{B} to \textbf{D}) Schematic (above) and corresponding LEED pattern (below) for the ARRES measurement for (A)  $\vec{k}_\parallel$ = 0, (B) $\vec{k}_\parallel\,\ne \,0$, and (C) $\vec{k}_\parallel\,\ne \,0$ with a realignment of the Ewald sphere onto the center of the camera. Green circle and orange hexagon indicate edge of Ewald sphere and BZ of $\mathrm{CoNb_4Se_8}$. 
\label{fig:arres_technique}}
\end{figure}

Angle Resolved Reflection Electron Spectroscopy measurements were conducted using the QSPLEEM at the Molecular Foundry at a temperature of 25~K. 

Details of this technique are presented in Fig.~\ref{fig:arres_technique}, and build upon Refs.~\cite{Jobst2015,Jobst2016}. As depicted in Fig.~\ref{fig:arres_technique}A the QSPLEEM is equipped with illumination and imaging equalizers which shift electron trajectories into and out of the beam splitter, respectively.  Slight deflections of the electrons $d\vec{x}_1$ (indicated by grey arrows) upstream of the beam splitter adjusts the particular electron trajectory taken through the microscope, effectively changing the parallel momentum $\vec{k}_\parallel$ of the electrons impinging upon and reflecting off of the sample, respectively. . As mentioned in the main text, the incident $\vec{k}_\parallel$ of electrons that are absorbed into the sample correspond to unoccupied electronic states at that same momentum. For an incident electron with energy $E$, the parallel momentum is defined by $\vec{k}_\parallel = E \sin \theta_{inc}$ where $\theta_{inc}$ is the angle of incidence on the sample controllable by the illumination equalizer. At normal incidence (Fig.~\ref{fig:arres_technique}B), for electron optics set to LEED mode the detector displays a low energy diffraction pattern corresponding to the order on the sample surface. Upon tuning $\theta_{inc}$,  nonzero incident momentum $\vec{k}_\parallel\ne0$  (Fig.~\ref{fig:arres_technique}C) causes the sampled diffraction pattern within the Ewald sphere (green circle) to shift to a new center about  the reflected angle angle $\theta_{ref}$. The imaging equalizer deflects reflected electrons by $d\vec{x}_2$ downstream of the beam splitter such that the specular spot centered along an off-center trajectory returns to the center of the detector (Fig.~\ref{fig:arres_technique}D), whereby a measurement of the intensity at detector center corresponds to the sample electron reflectivity at incident momentum  $\vec{k}_\parallel\ne0$. This procedure can be repeated for every energy, momentum, and spin value in the user's choice coordinate array. 

\textit{Linear Mapping between Pixel Space and Equalizer Space:}
To establish a mapping between pixel coordinates on a detector and the equalizer values required to position an electron beam, we assume a linear relationship between pixel position $\mathbf{P}$ and equalizer values $\mathbf{X}$:

\begin{equation}
\mathbf{P} = \mathbf{\Omega} \cdot \mathbf{X}
\end{equation}

where:
\begin{align}
\mathbf{P} &= \begin{bmatrix} P_x \\ P_y \end{bmatrix} \quad \text{(pixel position relative to center)} \\
\mathbf{X} &= \begin{bmatrix} X_x \\ X_y \end{bmatrix} \quad \text{(equalizer values in mA)} \\
\mathbf{\Omega} &= \begin{bmatrix} \alpha & \beta \\ \gamma & \delta \end{bmatrix} \quad \text{(transformation matrix)}
\end{align}

Using two known calibration points, we have:
\begin{align}
\text{Point 0:} \quad \begin{bmatrix} P_{0x} \\ P_{0y} \end{bmatrix} &= \mathbf{\Omega} \begin{bmatrix} X_{0x} \\ X_{0y} \end{bmatrix} \\
\text{Point 1:} \quad \begin{bmatrix} P_{1x} \\ P_{1y} \end{bmatrix} &= \mathbf{\Omega} \begin{bmatrix} X_{1x} \\ X_{1y} \end{bmatrix}
\end{align}

This can be written in matrix form:
\begin{equation}
\begin{bmatrix} P_{0x} & P_{1x} \\ P_{0y} & P_{1y} \end{bmatrix} = 
\begin{bmatrix} \alpha & \beta \\ \gamma & \delta \end{bmatrix}
\begin{bmatrix} X_{0x} & X_{1x} \\ X_{0y} & X_{1y} \end{bmatrix}
\end{equation}

or more compactly:
\begin{equation}
\mathbf{P}_{\text{cal}} = \mathbf{\Omega} \cdot \mathbf{X}_{\text{cal}}
\end{equation}

The solution to this equation can be found using Cramer's rule. For $D = \det(\mathbf{X}_{\text{cal}})$, the transformation matrix elements are:
\begin{align}
\alpha &= -\frac{X_{1y} P_{0x} - X_{0y} P_{1x}}{D} \\
\beta &= \frac{X_{1x} P_{0x} - X_{0x} P_{1x}}{D} \\
\gamma &= -\frac{X_{1y} P_{0y} - X_{0y} P_{1y}}{D} \\
\delta &= \frac{X_{1x} P_{0y} - X_{0x} P_{1y}}{D}
\end{align}

To find the equalizer values for an arbitrary pixel position, we need: $\mathbf{X} = \mathbf{\Omega}^{-1} \cdot \mathbf{P}$ 
Where  $\mathbf{\Omega}^{-1} = \frac{1}{\det(\mathbf{\Omega})} 
\begin{bmatrix} \delta & -\beta \\ -\gamma & \alpha \end{bmatrix}$ .

\textit{Polarization Measurements:}
As mentioned in the main text, spin-polarized ARRES measurements are conducted by measuring the reflectance contrast to an electron beam with spin polarization $P$ (in our case, $P \approx 0.3$). The spin polarization of the unoccupied electronic states are found by $A=\frac{1}{P} \frac{I_{+}-I_{-}}{I_{+}+I_{-}}$, where $I_+$ ( $I_-$) is the reflected intensities for spins polarized parallel (antiparallel) to a spin vector ($\hat{s}$ )\cite{rougemaille_magnetic_2010,delaFiguera2013}. In all measurements for the main text, we set $\hat{s}$ to be along the $c$ axis of the sample, parallel to the sample Néel vector.  Prior to measurement, the spin vector $\hat{s}$ is calibrated following the procedure in Ref.~\cite{rougemaille_self-organization_2006} . 

\begin{figure}
\centering \includegraphics[width=0.7\textwidth]{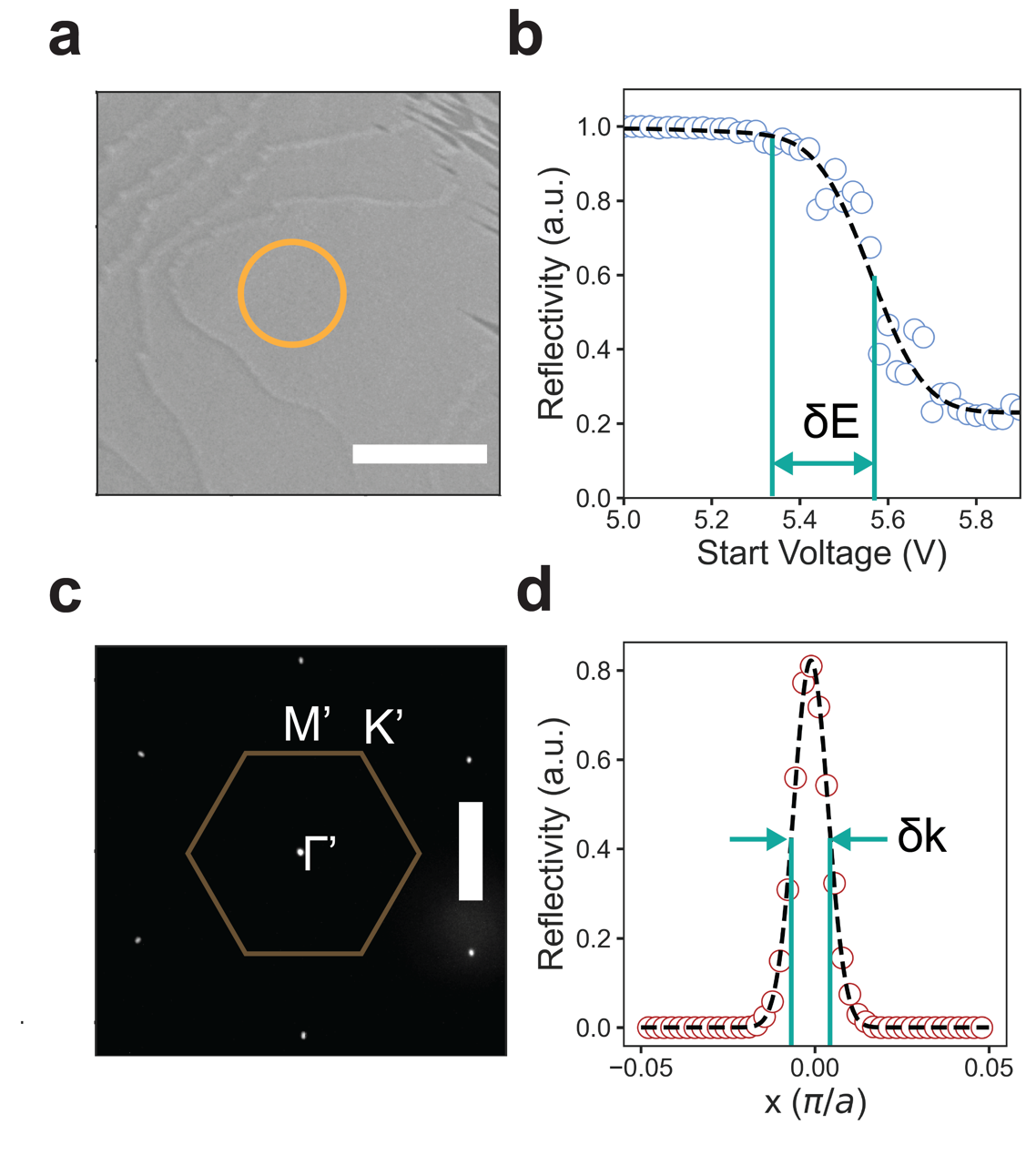}
\caption{\textbf{Energy and Momentum Resolution in SPLEEM} \textbf{(A)} LEEM image of $\mathrm{CoNb_4Se_8}$ at 25~K for start voltage of 5V. Scale bar: $10\,\mu$m. \textbf{(B)} LEEM reflectivity versus incident electron start voltage associated with region of sample in A. Dashed black curve indicates Fermi-Dirac fit described in the text. \textbf{(C)} LEED pattern for electron start voltage of 48V.  Scale bar: $\pi/2a $.  \textbf{(D)} Lineout along (0,0) spot in C. Dashed black curve indicates Gaussian peak fit to the data}
\label{fig:spleem_res}
\end{figure}
\textit{SPLEEM Resolution: }
The overall experimental resolution ($\delta E =$225 meV, $\delta k\,=\,0.01 \,\pi/a$)  is described  following Fig. \ref{fig:spleem_res}.  Samples become electron transparent for incident electron energy  ($E$)  above the work function ($\Phi$)\cite{delaFiguera2013}.  Given a certain region of interest on your sample ($\mathbf{r}$), depicted via LEEM mode in Fig. \ref{fig:spleem_res}A, the energy resolution can be determined from the width of an error function fit to the intensity $I(\mathbf{r},E)$ of that region as a function of incident electron start voltage ($E$): 
\begin{equation}
    f(E) = \frac{1}{2}\left[1 - \text{erf}\left(\frac{E-\Phi(\mathbf{r})}{\sqrt{2(\sigma^2+(k_BT)^2)}}\right)\right]
\end{equation}
Here $\sigma$ denotes the standard deviation of a gaussian which is convolved with the Fermi-Dirac distribution of electrons at temperature $T$ . For low temperature ($k_BT<< \sigma $), the energy resolution is the FWHM of that gaussian, i.e.  $\delta E = 2 \sigma \sqrt{2 \ln{2}} = 225$ meV.

Similarly, the momentum resoution can be determined from LEED data, shown in Fig. \ref{fig:spleem_res}C for the sample measured in the main text. Momentum resolution $\delta \vec{k}/|\vec{G}|$ is found as the ratio between the width of the diffraction spots to the BZ size. Here, the width of the LEED spots are the FWHM extracted from Gaussian fit to the (0,0) peak in LEED data (Fig. \ref{fig:spleem_res}D), whereas the distance between the spots corresponds to the reciprocal lattice vector with norm given $|\vec{G}|$.  Combining these we find an overall momentum resolution of $0.01 \,\pi/a$ .

\subsection*{Computational Methods}
Our density functional theory (DFT) calculations use the projector augmented wave (PAW) method ~\cite{blochlProjectorAugmentedwaveMethod1994} as implemented in the Vienna Ab Initio Simulation Package (VASP)~\cite{kresseEfficiencyAbinitioTotal1996b,kresseEfficientIterativeSchemes1996a,kresseInitioMolecularDynamics1993,kresseInitioMoleculardynamicsSimulation1994,kresseUltrasoftPseudopotentialsProjector1999} version 6.4.3 and the \texttt{potpaw.64} PAW dataset. All the calculations in the main text use the PBE functional~\cite{PBE}, without any dispersion or Hubbard corrections. We set the wavefunction energy cutoff at 900 eV and BZ integrations were performed on a $\Gamma$-centered $11\times11\times7$ $k$-grid with 100 meV of second-order Methfessel-Paxton smearing~\cite{methfesselHighprecisionSamplingBrillouinzone1989}. All calculations used an energy convergence criterion of $10^{-7}$ eV, and a force convergence criterion of $10^{-3} \text{eV}/$\AA.  Band structures and 3D constant energy surfaces (Figs. \ref{fig:crystalstructure}, \ref{fig:arpes_mgm}, \ref{fig:arpes_kgk} and \ref{fig:kzdep}) were plotted using modified versions of \texttt{sumo}~\cite{sumo} and \texttt{IFermi}~\cite{ifermi}, respectively. The Mulliken gross populations were calculated using LOBSTER~\cite{lobster, lobster2}. 

To select the optimal DFT functional, benchmarked the performance of several exchange-correlation functionals, van-der-Waals corrections, and Hubbard U corrections, as shown in Table \ref{tab:funcs}. These calculations used a $9\times9\times5$ $k$-grid and an 800~eV cutoff. As seen from the table, the PBE functional without corrections performs nearly as well as PBE corrected by a Hubbard $U$ of 0.75~eV and a D3 van-der-Waals correction with Becke-Johnson damping. For simplicity, we choose the uncorrected PBE for our calculations.

\begin{table}[!ht]
    \centering
    \caption{Comparison of exchange-correlation functionals for structural (lattice constants $a$ and $c$) and magnetic (magnetic moment $m$ per Co) parameters in CoNb$_4$Se$_8$. Hubbard $U$ values are in eV and we used Dudarev's approach~\cite{dudarev}. The magnetic moment values are computed directly from the DFT, and are slightly smaller than the more accurate values obtained via Mulliken population analysis used in the main text. \label{tab:funcs}}
    \begin{tabular}{l|lll|lll}
        ~ & $a$ Error (\%) & $c$ Error (\%) & $M$ error (\%) & $a$ (\AA) & $c$ (\AA) & $m$ ($\mu_B$)  \\ \hline
        Experiment (5K)~\cite{regmi2024altermag_arxiv} & $-$ & $-$ & $-$ & 6.904 & 12.321 & 1.375  \\ \hline
        PBE\cite{PBE}  & +0.860 & +1.251 & +6.256 & 6.964 & 12.475 & 1.461  \\ 
        PBE\cite{PBE}-D3(BJ)\cite{DFT-D3,DFT-D3BJ} & -0.818 & -2.066 & -50.327 & 6.848 & 12.066 & 0.683  \\
        PBE\cite{PBE}-D3(BJ)\cite{DFT-D3,DFT-D3BJ}, $U$=0.5 & -0.606 & -1.659 & -11.417 & 6.862 & 12.117 & 1.218  \\ 
        PBE\cite{PBE}-D3(BJ)\cite{DFT-D3,DFT-D3BJ}, $U$=0.75 & -0.548 & -1.341 & +2.256 & 6.866 & 12.156 & 1.406  \\ 
        PBE\cite{PBE}-D3(BJ)\cite{DFT-D3,DFT-D3BJ}, $U$=1 & -0.492 & -1.192 & +11.856 & 6.870 & 12.174 & 1.538  \\ 
        PBE\cite{PBE}-D3(BJ)\cite{DFT-D3,DFT-D3BJ}, $U$=3 & -0.224 & +0.368 & +53.893 & 6.889 & 12.366 & 2.116  \\ 
        PBEsol\cite{PBEsol} & -0.926 & -2.045 & -63.490 & 6.840 & 12.069 & 0.502 \\ 
        PBEsol\cite{PBEsol}-D3(BJ)\cite{DFT-D3,DFT-D3BJ} & -2.326 & -3.910 & $-$ & 6.744 & 11.840 & 0.000  \\ 
        R$^2$SCAN\cite{r2SCAN} & +1.038 & +2.256 & +49.166 & 6.976 & 12.599 & 2.051  \\ 
    \end{tabular}
\end{table}

For the slab calculations, we constructed a symmetric slab from the relaxed structure, with 9 layers of NbSe$_2$ and 8 Co ions, with a NbSe$_2$ termination. These calculations employ Quantum ESPRESSO \cite{} v7.3 with the fully relativistic ONCVPSP pseudopotentials, and a planewave energy cutoff of 100 Ry. Spin-orbit coupling was included self-consistently in all slab calculations.

\begin{figure}
    \centering
    \includegraphics[width=0.6\linewidth]{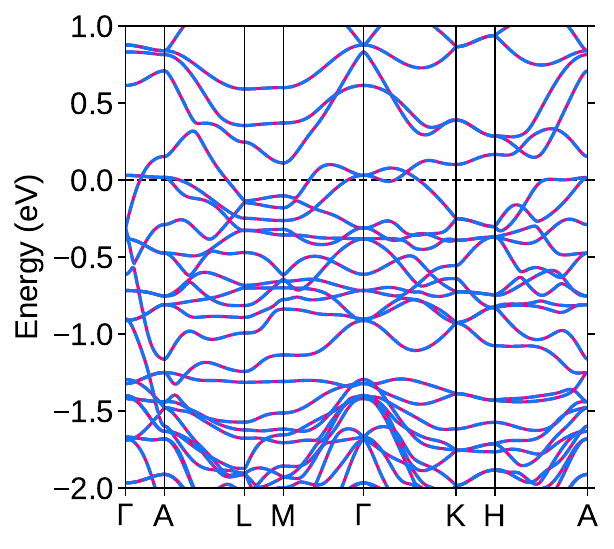}
    \caption{\label{fig:full_bands} The DFT-calculated spin-polarized band structure of CoNb$_4$Se$_8$ along high symmetry lines in the BZ without SOC. Spin-up (solid pink) and spin-down (dashed blue) bands are degenerate along every high symmetry line. The Fermi level is set to 0~eV and is marked by the dashed line.}
\end{figure}

\begin{figure}
    \includegraphics[width=1.\linewidth]{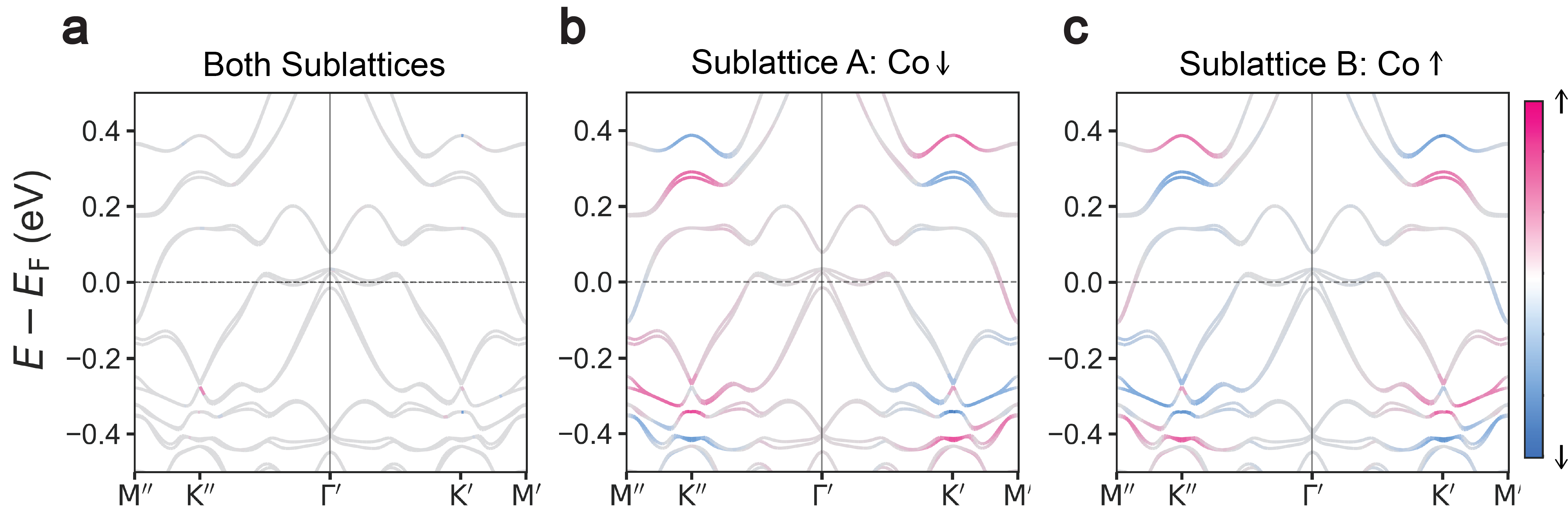}
    \caption{\label{fig:slab_sublattice} \textbf{(A-C) }The DFT-calculated spin-polarized band structure of CoNb$_4$Se$_8$ for both sublattices (\textbf{A-}), and for each individual Co sublattice (\textbf{B-,C-}) for spin polarized along the $c$ axis. The Fermi level is set to 0~eV and is marked by the dotted line.}
\end{figure}
\subsection*{Supplementary Text}

\subsubsection*{Tight-binding model}

\begin{figure}
    \includegraphics[width=1.\linewidth]{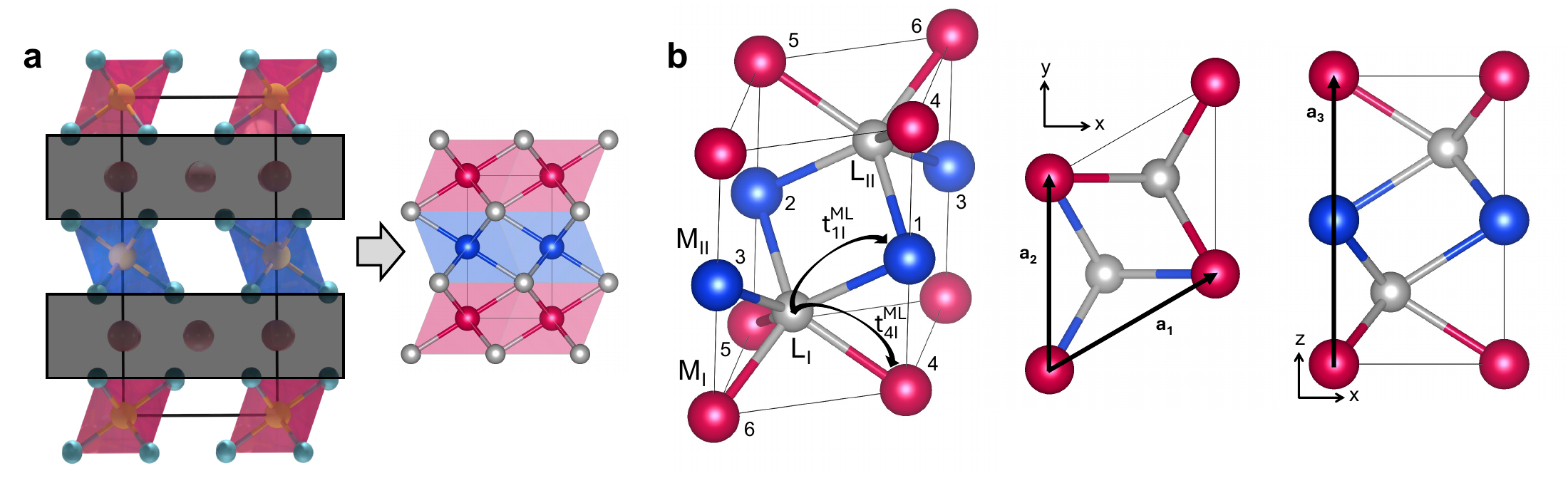}
    \caption{(\textbf{A}) Mapping of the CoNb$_4$Se$_8$ lattice into a simpler lattice by removing the Nb atoms and some of the Se atoms. (\textbf{B}) Unit cell of the tight-binding model, with the sublattices, lattice vectors and coordinate axes identified. Numbers $i=1-6$ indicate the neighbors of each ligand site $j=\text{I, II}$ used to define the hoppings, denoted by $t_{ij}^{ML}$.}
    \label{fig_TBlattice}
\end{figure}

To formulate a tight-binding model for CoNb$_4$Se$_8$, we follow the microscopic understanding from Ref. \cite{Vila2024}, which includes orbital, sublattice and spin degrees of freedom. The single-particle Hamiltonian takes the form
\begin{equation}\label{eq_H}
\hat{\mathcal{H}} = \hat{\mathcal{H}}_0 + \hat{\mathcal{H}}_{CF} + \hat{\mathcal{H}}_{ex},
\end{equation}
with $\hat{\mathcal{H}}_0$ containing the hopping integrals, $\hat{\mathcal{H}}_{CF}$ the crystal field and $\hat{\mathcal{H}}_{ex}$ the mean-field exchange interaction. Regarding the real-space lattice, we note that the lattice of CoNb$_4$Se$_8$ can be simplified by removing Nb and some Se atoms, while keeping the same space group, as depicted in Fig. \ref{fig_TBlattice}A. In fact, the resulting lattice corresponds to the actual lattice structure of other \textit{g}-wave altermagnets such as MnTe or CrSb~\cite{smejkal_beyond_2022}.

The unit cell is shown in Fig. \ref{fig_TBlattice}B, and includes two magnetic (M) sites and two ligand (L) sites. The fractional coordinates $\textbf{r}=(x,y,z)$ read
\begin{align*}
    \text{M}_\text{I} &= (0,0,0) \\
    \text{M}_\text{II} &= (0,0,1/2) \\
    \text{L}_\text{I} &= (1/4,1/4,1/4) \\
    \text{L}_\text{II} &= (3/4,3/4,3/4).
\end{align*}

The terms in Eq. \ref{eq_H} are defined as:
\begin{align}
    \hat{\mathcal{H}}_{0} &= \sum_{i, \tau} c_{i\tau}^\dagger \varepsilon^M_{i\tau} c_{i\tau} + \sum_{i, \tau} b_{i\tau}^\dagger \varepsilon^L_{i\tau} b_{i\tau} + \sum_{\langle i,j \rangle, \tau, \tau^\prime} c_{i\tau}^\dagger [t_{ij}^{ML}]_{\tau\tau^\prime} b_{j\tau^\prime} + H.C  \label{eq_H0} \\
    \hat{\mathcal{H}}_{CF} &= \sum_{i,\tau, \tau^\prime} c_{i\tau}^\dagger [\Delta_i]_{\tau\tau^\prime} c_{i\tau^\prime} \label{eq_HCF} \\
    \hat{\mathcal{H}}_{ex} &= m_{ex} \sum_{i,s,s^\prime} c_{is}^\dagger [\textbf{m}_i \cdot \textbf{s}]_{ss^\prime} c_{is^\prime}.
\end{align}
Here, $c^\dagger$ ($b^\dagger$) and $c$ ($b$) are creation and annihilation operators acting on the M (L) site, respectively. Sites are labeled by the $i$ and $j$ indices, $\langle \rangle$ denotes nearest-neighbor interaction and $\tau$ and $s$ are orbital and spin indices, respectively. 

Next, let us define in more detail each term of the Hamiltonian. Starting from the exchange interaction, which denotes the coupling of electrons with local magnetic moments $\textbf{m}_i$ with strength $m_{ex}$, we take $\textbf{m} = + (0,0,\hat{\textbf{z}})$ for sublattice I  and $\textbf{m} = - (0,0,\hat{\textbf{z}})$ for sublattice II, as this corresponds to the magnetic order in CoNb$_4$Se$_8$. 

The electronic term $\hat{\mathcal{H}}_0$ is formed by the onsite energies of the M and L sites and by the hoppings between said sites. To define the hopping terms, we resort to the tabulated Slater-Koster overlap integrals between two pair of orbitals at neighboring sites $i$ and $j$~\cite{Slater1954}. In this way, given a L site and its six M neighbors (Fig. \ref{fig_TBlattice}B), one can define six different hoppings. The simplest and most efficient way is to use the Slater-Koster parametrization to define only one of those hoppings, and then obtain the rest by applying symmetry operations. For example, we define the hopping $t_{ij}^{ML} = t_{1I}^{ML}$ as the hopping from the L site in sublattice I to the M site labeled as 1 in Fig. \ref{fig_TBlattice}B. For this minimal model, we consider the five cubic $d$ orbitals for the M sites and choose a single $p_z$ orbital for the L sites for concreteness and simplicity. Hence,
\begin{equation}
    |\Psi \rangle = 
    \left(
p_z^\text{I} \quad p_z^\text{II} \quad d_{xy}^\text{I} \quad d_{x^2-y^2}^\text{I} \quad d_{xz}^\text{I} \quad d_{yz}^\text{I} \quad d_{z^2}^\text{I} \quad d_{xy}^\text{II} \quad d_{x^2-y^2}^\text{II} \quad d_{xz}^\text{II} \quad d_{yz}^\text{II} \quad d_{z^2}^\text{II} 
    \right)^T 
    \otimes 
    \begin{pmatrix}
    \uparrow \\ \downarrow
    \end{pmatrix},
\end{equation}
and
\begin{align}
    t_{1I}^{ML} &= \begin{pmatrix} \langle d_{xy}^\text{II}| \hat{\mathcal{H}}_0 | p_z^\text{I} \rangle  & \langle d_{x^2-y^2}^\text{II}| \hat{\mathcal{H}}_0 | p_z^\text{I} \rangle & \langle d_{xz}^\text{II}| \hat{\mathcal{H}}_0 | p_z^\text{I} \rangle & \langle d_{yz}^\text{II}| \hat{\mathcal{H}}_0 | p_z^\text{I} \rangle & \langle d_{z^2}^\text{II}| \hat{\mathcal{H}}_0 | p_z^\text{I} \rangle \end{pmatrix}^T \nonumber \\
    &= \begin{pmatrix} 0 & \frac{\sqrt{3}}{2} l^2 n V_\sigma - l^2 n V_\pi & \sqrt{3} l n^2 V_\sigma + l(1-2n^2)V_\pi& 0 & n(n^2-\frac{l^2}{2})V_\sigma + \sqrt{3} l^2 n V_\pi  \end{pmatrix}^T \otimes 
    \begin{pmatrix}
    \uparrow \\ \downarrow
    \end{pmatrix},
\end{align}
where $V_\pi$ and $V_\sigma$ are the parametrizations of the $\pi$ and $\sigma$ chemical bonds, and $\alpha$, $\beta$ and $\gamma$ are the direction cosines for the $x$, $y$ and $z$ Cartesian directions, respectively, of the $t_{1I}^{ML}$ hopping (e.g. $\alpha$ is related to the angle between $t_{1I}^{ML}$ and the $x$ axis). Having defined this hopping term, the rest are straightforwardly obtained as
\begin{align}
    t_{2I}^{ML} &= t_{1I}^{ML} {C_{3z}} ,\;\;
    t_{3I}^{ML} = t_{2I}^{ML} {C_{3z}} ,\;\;
    t_{4I}^{ML} = t_{1I}^{ML} {M_z} ,\;\;
    t_{5I}^{ML} = t_{2I}^{ML} {M_z} ,\;\;
    t_{6I}^{ML} = t_{3I}^{ML} {M_z} , \\
    t_{1II}^{ML} &= t_{6I}^{ML} {C_{6z}} ,\;\;
    t_{2II}^{ML} = t_{5I}^{ML} {C_{6z}} ,\;\;
    t_{3II}^{ML} = t_{4I}^{ML} {C_{6z}} ,\;\;
    t_{4II}^{ML} = t_{3I}^{ML} {C_{6z}} ,\;\;
    t_{5II}^{ML} = t_{2I}^{ML} {C_{6z}} ,\;\;
    t_{6II}^{ML} = t_{1I}^{ML} {C_{6z}} ,\;\; \nonumber
\end{align}
where $C_{3z}$ ($C_{6z}$) is a three-fold (six-fold) rotation about the $z$ axis. Regarding the onsite energies, $\varepsilon^M_{i\tau}$ encodes the energies of each $d$ orbital, while $\varepsilon^L_{i\tau}$ describes the energy of the $p_z$ orbitals. We note that while other models require of third neighbor hoppings between M sites to produce NRSS \cite{Yang_et_al:2024}, our model captures NRSS at the nearest neighbor hopping between M and L sites.

To model the crystal field term, we note that the point group of the M sites is $D_{3d}$. The corresponding crystal field is usually referred to in the literature as trigonally distorted octahedra, where the three-fold degenerate $t_{2g}$ orbitals split into a single $a_{1g}$ state and two degenerate $e_g$ states. The Hamiltonian can then be easily written in diagonal form in the trigonal basis. However, the change of basis from trigonal to cubic orbitals is not unique, and several conventions have been used \cite{Oreilly1971, Perumareddi1973, Winter2017}. To avoid discrepancies, we instead derive the $D_{3d}$ crystal field from spherical tensors, which allow us to express the Hamiltonian in the basis of spherical harmonics which has a straightforward transformation to cubic \textit{d} orbitals. In the spherical tensor basis, a general Hamiltonian can be written as:
\begin{equation}
\hat{h} = \sum_{k=0}^{2l} \sum_{q=-k}^k h_q^k \hat{T}_q^k,
\end{equation}
where $h_q^k$ is a component and $\hat{T}_q^k$ the basis vector. When working with \textit{d} orbitals, $l=2$ and $k$ runs up to index 4. By using the symmetries of the $D_{3d}$ point group, many coefficients become zero and the resulting Hamiltonian becomes:
\begin{equation}\label{eq_CFA}
\hat{\mathcal{H}}_{CF}^{D_{3d}} =  h_0^0 \hat{T}_0^0 + h_0^2 \hat{T}_0^2 + h_0^4 \hat{T}_0^4 + h_3^4 \left( \hat{T}_3^4 - \hat{T}_{-3}^4 \right),
\end{equation}
that is, there are 4 independent parameters, all being real. Eq. \ref{eq_CFA} is obtained by considering the crystalline environment of $\text{M}_\text{I}$, where one of the $C_{2}$ rotation axes is parallel to the $y$ axis. By applying a $6^\pm_{001}$ rotation to $\hat{\mathcal{H}}_{CF}^{D_{3d}}$, the crystal field term for sublattice $\text{M}_\text{II}$ is obtained. Thus:
\begin{align}
\hat{\mathcal{H}}_{CF}^{I} &= h_0^0 \hat{T}_0^0 + h_0^2 \hat{T}_0^2 + h_0^4 \hat{T}_0^4 +  h_3^4 \left( \hat{T}_3^4 - \hat{T}_{-3}^4 \right) \\
\hat{\mathcal{H}}_{CF}^{II} &= h_0^0 \hat{T}_0^0 + h_0^2 \hat{T}_0^2 + h_0^4 \hat{T}_0^4 - h_3^4 \left( \hat{T}_3^4 - \hat{T}_{-3}^4 \right).
\end{align}
Interestingly, the crystal field in both sites is the same except the term $h_3^4$ that changes sign. To express the above terms in the spherical harmonic basis $|l,m\rangle \langle l',m|$, we rely on the Wigner-Eckart theorem. In this way, and considering $l'=l$, the matrix elements between both basis are related as:
\begin{equation}
    \langle l,m | \hat{\mathcal{H}}_{CF}^{D_{3d}} | l,m'\rangle = \sum_{k=0}^{2l} \sum_{q=-k}^k \mu_{mm'}^{lkq} h_q^k,
\end{equation}
with
\begin{equation}
    \mu_{mm'}^{lkq} = (-1)^{l-m} \sqrt{2k + 1} \begin{pmatrix}
        l & k & l \\
        -m & q & m'
    \end{pmatrix},
\end{equation}
where the last term is the 3j symbol \cite{Newman_Ng_2000}. In this manner, the 5x5 matrix representation in spherical harmonics (with basis order $|2,-2\rangle,|2,-1\rangle,|2,0\rangle,|2,1\rangle,|2,2\rangle$) is 
\begin{equation}
\hat{\mathcal{H}}_{CF,|l,m\rangle}^{I/II} = \begin{pmatrix}
    D_2 & 0 & 0 & \pm \lambda & 0 \\
    0 & D_1 & 0 & 0 & \mp \lambda \\   
    0 & 0 & D_0 & 0 & 0 \\
    \pm \lambda & 0 & 0 & D_1 & 0 \\
    0 & \mp \lambda & 0 & 0 & D_2
    \end{pmatrix},
\end{equation}
where $D_2 = h_0^0-\frac{2}{7}h_0^2+\frac{1}{21}h_0^4$, $D_1 = h_0^0+\frac{1}{7}h_0^2-\frac{4}{21}h_0^4$, $D_0 = h_0^0+\frac{2}{7}h_0^2+\frac{2}{7}h_0^4$ and $\lambda = \frac{1}{3}\sqrt{\frac{5}{7}}h^4_3$. Finally, we can transform the above matrix into a basis of cubic \textit{d} orbitals and obtain
\begin{equation}
\hat{\mathcal{H}}_{CF,d}^{I/II} =  \begin{pmatrix}
    D_2 & 0 & 0 & \pm \lambda & 0 \\
    0 & D_2 & \mp \lambda & 0 & 0 \\   
    0 & \mp \lambda & D_1 & 0 & 0 \\
    \pm \lambda & 0 & 0 & D_1 & 0 \\
    0 & 0 & 0 & 0 & D_0
    \end{pmatrix},
\end{equation}
where each term corresponds to $\Delta_i$ in Eq. \ref{eq_HCF}.

\begin{figure}
    \centering \includegraphics[width=0.7\linewidth]{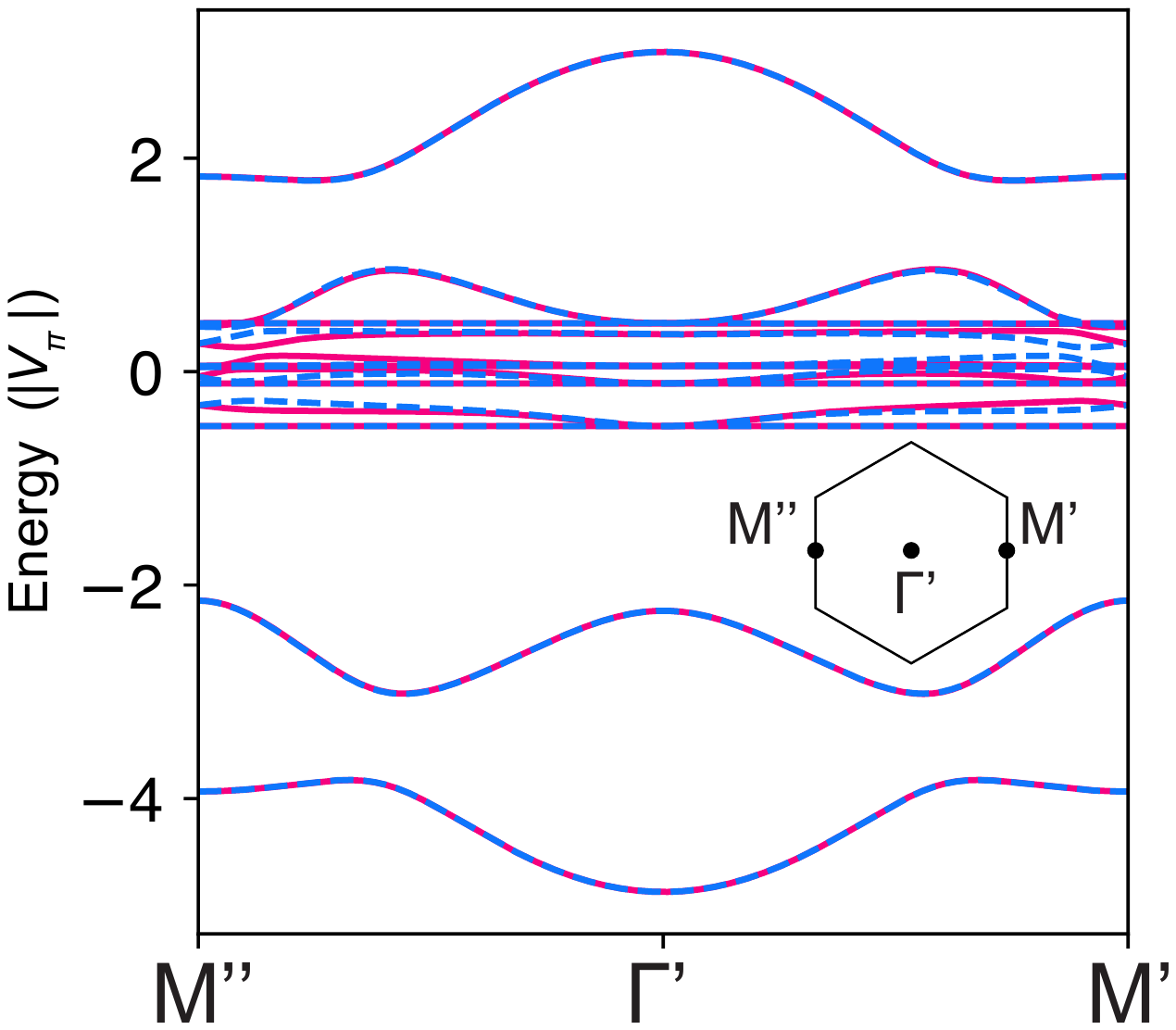}
    \caption{Band structure of Eq. \eqref{eq_H} using the parameters described in the text. Solid pink (dashed blue) bands correspond to spin up (spin down) and the inset shows the Brillouin zone and high-symmetry points.}
    \label{fig_TB_bands}
\end{figure}

With each Hamiltonian term understood, one can Fourier transform Eq. \eqref{eq_H} to plot the band structure as shown in the main text. We plot in Fig. \ref{fig_TB_bands} the band structure of Eq. \eqref{eq_H} for up and down spins. We use the following parameters, which are the same used for Fig. 2 in the main text, but here we show all bands at all energies rather than a narrower energy window. We take $V_\pi=1$ as a reference value and use $|V_\pi|$ as a unit. Then we take $V_\sigma = 0.5$, $m_{ex} = 0.2$, $h^0_0 = 0$, $h^2_0 = 0.2$, $h^4_0 = 0.2$ and $h^4_3 = 1$. Since the simplified tight-binding lattice in Fig. \ref{fig_TBlattice}A is isostructural to CrSb, where the angle between $t_{1I}^{ML}$ and the $x$ axis is approximately 32\degree, we use $\alpha=\cos(32\frac{\pi}{180})$, $\gamma=\cos(58\frac{\pi}{180})$. Finally, we set the onsite energies of all $d$ orbitals to zero, $\varepsilon^M_{i\tau} = 0 \; \forall \; i,\tau$, and the onsite energy for the ligand $p_z$ orbitals to $\varepsilon^L_{i\tau} = -2 \; \forall \; i,\tau$.

\subsection*{Supplementary Figures}

\subsubsection*{3D electronic structure of CoNb$_4$Se$_8$}

\begin{figure}
\includegraphics[width=1.\textwidth]{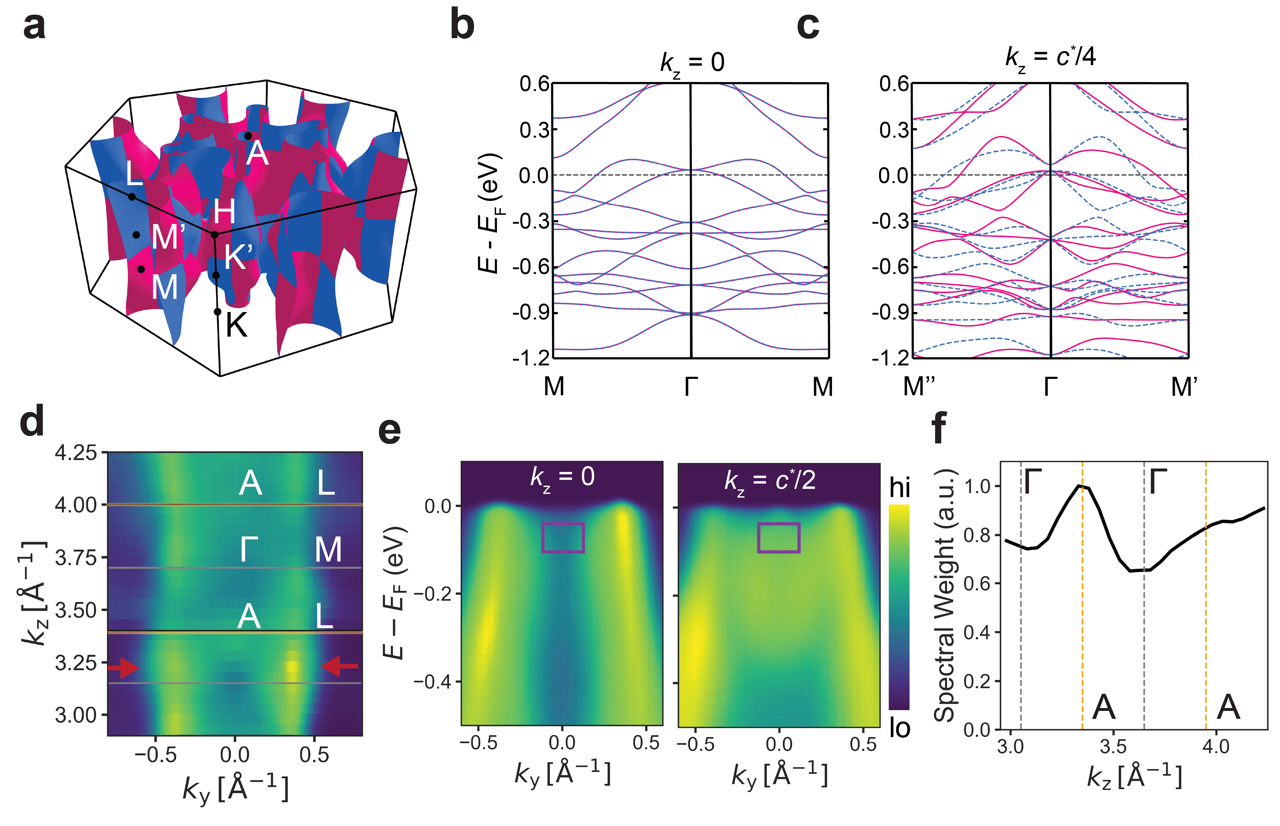}
\caption{\textbf{3D Electronic Structure of $\mathrm{CoNb_4Se_8}$ } \textbf{(A)} Fermi surface of  $\mathrm{CoNb_4Se_8}$ , calculated using spin-polarized DFT \textbf{(B, C)}. Calculated spin-polarized band structure along $\mathrm{\Gamma - M - A -L}$ plane for $k_\mathrm{z} =0$ (B) and $k_\mathrm{z} = c^*/4 = \pi/(2c)$ (C). Spin-up (-down) bands are shown in solid pink (dashed blue) and the Fermi level is set to 0~eV and marked with a dashed link. \textbf{(D)}  ARPES Fermi surface cuts along  $\mathrm{\Gamma - M - A -L}$  plane. Red arrows indicate $k_\mathrm{z}$ value for photon energy of 55 eV. \textbf{(E)} corresponding spectra at $k_\mathrm{z} =0$ (left) and $k_\mathrm{z} = c^*/2 $ (right). \textbf{(F)} Spectral weight as a function of $k_\mathrm{z}$ , extracted from region indicated by the purple rectangle in \textbf{D}.}
\label{fig:kzdep}
\end{figure}

Here in Fig. \ref{fig:kzdep} we demonstrate that altermagnetic spin-splitting occurs away from time-reversal invariant momenta (TRIMs). Along high symmetry directions, i.e. $\Gamma - \mathrm{K}$ and $\Gamma - \mathrm{M}$, the spin-splitting in the Fermi surface (Fig. \ref{fig:kzdep}A) changes sign, implying a zero crossing at these TRIMs. More explicitly, the DFT band structure along $k_z = 0$ (Fig. \ref{fig:kzdep}B) has spin degeneracy, whereas at $k_z = c^*/4$ (Fig. \ref{fig:kzdep}C), there is significant spin-splitting, confirming previous predictions\cite{regmi2024altermag_arxiv}. 

The experimental 3D electronic structure of $\mathrm{CoNb_4Se_8}$ as measured in ARPES at 13K is presented in Fig. \ref{fig:kzdep}D.  Along the BZ edge the ARPES spectra present maximal intensity at normal emission, whereas at BZ center, the spectra present minimal normal emission intensity.  This periodic structure, summarized in Fig.  \ref{fig:kzdep}F indicates a $c^* \approx 0.6$~\AA$^{-1}$ corresponding to a c axis length of 10.5 \AA,  which is within 20\% of the experimentally determined $c = 12.4$~\AA. All data in the main manuscript were taken at a photon energy of 55 eV, corresponding to an intermediate $k_z \approx c^*/4$, as indicated by the red arrows in panel D.  

As the large c axis in $\mathrm{CoNb_4Se_8}$ contributes a substantial amount of broadening along $k_z$, the photoemission spectra at a pparticular photon energy sample a range of $k_z$ values within the BZ\cite{devita2025opticalswitchinglayeredaltermagnet}. As such, we present in Fig. \ref{fig:arpes_mgm} a match between DFT and ARPES Fermi surfaces for $k_z = 0.18 c^*$, but reasonable agreement between calculation and experiment can be found within a range of $k_z = 0.2\pm0.1 c^*$. 

\subsubsection*{Suppression of NRSS above $T_\mathrm{N}$ along  $\mathrm{M^{\prime\prime} - \Gamma{^\prime} - M^{\prime}}$}
\begin{figure}
\includegraphics[width=1.\textwidth]{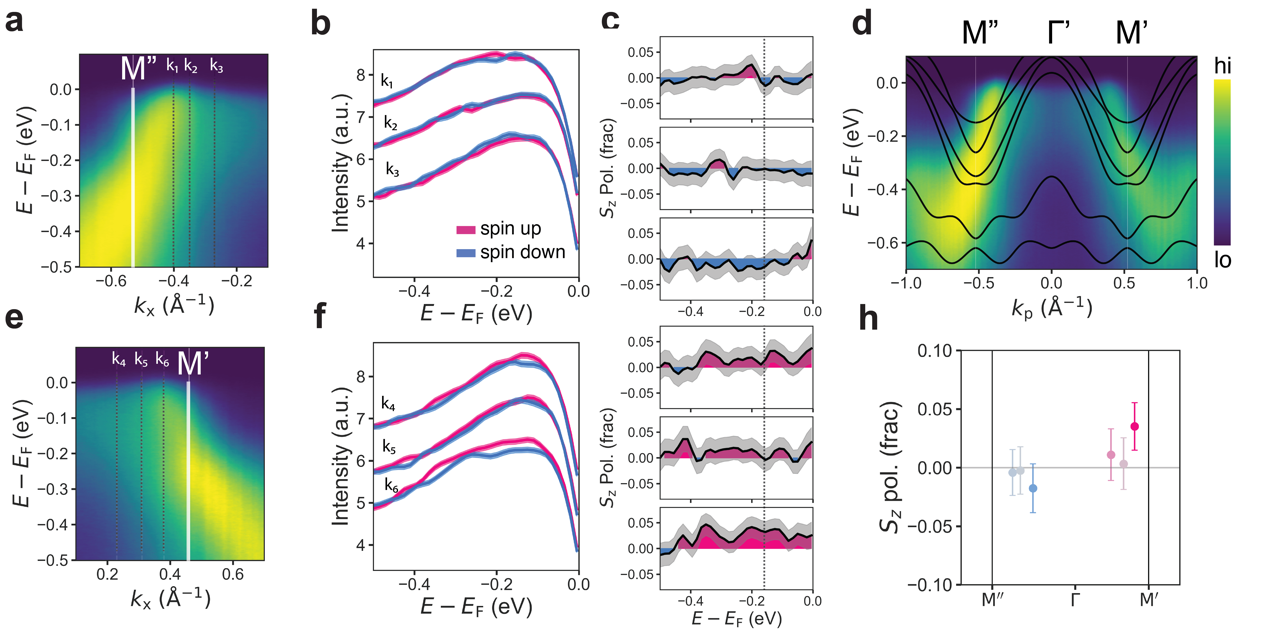}
\caption{\textbf{ARPES measurement of suppressed NRSS above $T_\mathrm{N}$ in CoNb$_4$Se$_8$.} \textbf{(A)} 200~K ARPES spectra along $\mathrm{M^{\prime\prime} - \Gamma{^\prime}}$ ,  \textbf{(B,C)} Spin-resolved EDCs spectra (\textbf{B-}) and corresponding spin polarization along the $z$ axis (C-})  at momenta indicated by black vertical lines in \textbf{A}.  Measurement uncertainty out to to $1\sigma$ outlined in grey. \textbf{(D)} Nonmagnetic DFT calculations of electronic structure along  $\mathrm{M^{\prime\prime} - \Gamma{^\prime} - M^{\prime}}$ incorporating both sublattices, overlaid onto ARPES data. \textbf{(E-G)} Same as (\textbf{A-C}) but for electronic structure along  $\mathrm{ \Gamma{^\prime} -M^{\prime}} \equiv C_{6z}^3(\mathrm{M^{\prime\prime} - \Gamma{^\prime}})$ direction in momentum space.\textbf{ (H)} $S_\mathrm{z}$ polarization as a function of momentum along $\mathrm{M^{\prime\prime} - \Gamma{^\prime} - M^{\prime}}$, extracted from binding energy depicted by dotted lines in (\textbf{C,G}).

\label{fig:arpes_mgm_aboveT_N}
\end{figure}

Angle-resolved photoemission spectroscopy (ARPES) measurements of $\mathrm{CoNb_4Se_8}$  (Fig.\ref{fig:arpes_mgm_aboveT_N}) capture electronic structure above  $T>T_\mathrm{N}$ . 

Recalling from the main text, prominent NRSS features included steep electron pockets originating from zone edge at $\approx E_
\mathrm{F}-200\,\mathrm{meV}$ and hole like bands extending from just above $E_
\mathrm{F}$ at zone center to $\approx E_
\mathrm{F}-300\,\mathrm{meV}$ at zone edge. In the nonmagnetic state, DFT predicts that the electron pockets at zone edge merge with the hole bands and lose their spin polarization, as shown in Fig. \ref{fig:arpes_mgm_aboveT_N}D.

Raw ARPES data along the $\mathrm{M^{\prime\prime} - \Gamma{^\prime}}$ and $\mathrm{\Gamma{^\prime} - M^{\prime}}$ directions in momentum space (Fig.\ref{fig:arpes_mgm_aboveT_N}A,E) confirm the existence of these features, with matrix elements still favoring the electronic pockets at the zone edge over the fainter but still visible hole-like bands which host the NRSS in the magnetic phase. 

Importantly, spin-resolved EDCs spectra along (Fig.\ref{fig:arpes_mgm_aboveT_N}C,G) confirm that hole-like features  $\approx$170 meV below $E_\mathrm{F}$ have a have zero spin polarization within 1 standard deviation $\sigma$, with a small residual polarization along $k_6$. These findings, summarized in  Fig.\ref{fig:arpes_mgm_aboveT_N}H, confirm the suppression of NRSS at $T>T_\mathrm{N}$.

\subsubsection*{Persistence of RSS above $T_\mathrm{N}$ along $\mathrm{K^{\prime\prime} - \Gamma{^\prime} - K^{\prime}}$} 
\begin{figure}
\includegraphics[width=1.\textwidth]{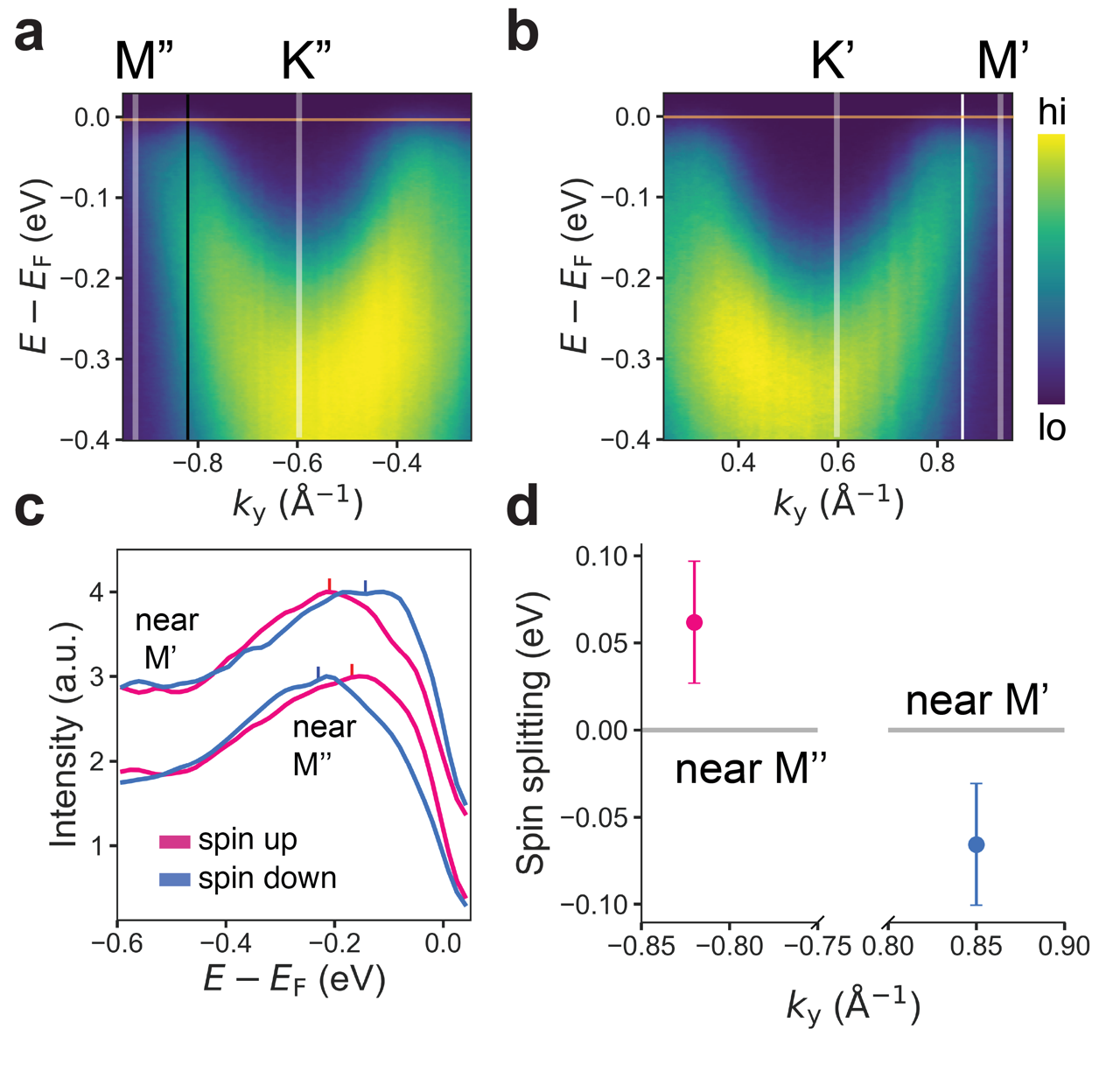}
\caption{\textbf{ARPES Measures Spin-Split Occupied Electronic Structure in CoNb$_4$Se$_8$ above $T_N$}. \textbf{(A,B)} 200K ARPES spectra along electron pockets surrounding $ \mathrm{K^{\prime\prime}} \equiv (6^\pm_{001})^3(\mathrm{K^{\prime}})$ (A) and $\mathrm{K^{\prime}}$ (B). (\textbf{C}) spin-resolved EDCs spectra along momenta indicated by black (white) vertical lines in A (B). Pink (blue) ticks indicate peak locations for bands polarized spin up (down). (\textbf{D}) Spin splitting, extracted from the difference in spin up and spin down band locations in \textbf{c,d}. 
}
\label{fig:spindep}
\end{figure}

Unlike NRSS, which is suppressed with temperature, the hallmark of relativistic spin splitting at inversion-broken surfaces is its persistence with temperature\cite{zhang_hidden_2014,bawden_spinvalley_2016,Gotlieb2018}. Here, we present 200K ARPES spectra of $\mathrm{CoNb_4Se_8}$  in (Fig.~\ref{fig:spindep}) along the momentum direction $\mathrm{M^{\prime\prime} - K^{\prime\prime} -\Gamma^\prime -K^{\prime} - M^{\prime}}$.  Similar to the low temperature data presented in the main text, the 200~K spectra (Fig.~\ref{fig:spindep}A, B) also exhibits electron pockets at the BZ corners with a similar depth and momentum splitting near $E_{\rm{F}}$, broadened by temperature effects. 

Above $T_\mathrm{N}$, energy distribution curves (EDCs) at momentum $k_1$ (Fig.~\ref{fig:spindep}C) still display spin polarization. States polarized along the +$c$ direction are shifted closer to the Fermi level by 60 ± 30 meV than those along $-c$. By contrast, at momentum $k_2 = 2^{\pm}_{001}(k_1) = (6^\pm_{001})^3(k_1)$, the spin splitting is inverted: states polarized along $+c$ are now 100 ± 30 meV further from $E_{\rm{F}}$. Fig.~\ref{fig:spindep}D summarizes this alternating spin splitting in $\mathrm{CoNb_4Se_8}$, which is within error bars of the magnitude seen in the altermagnetic phase, and consistent with the spin splitting which exists in the normal state of parent compound $\mathrm{NbSe_2}$ \cite{bawden_spinvalley_2016}. 


\subsubsection*{Collinear Antiferromagnetic Order in CoNb$_4$Se$_8$}

\begin{figure}
\centering\includegraphics[width=0.5\textwidth]{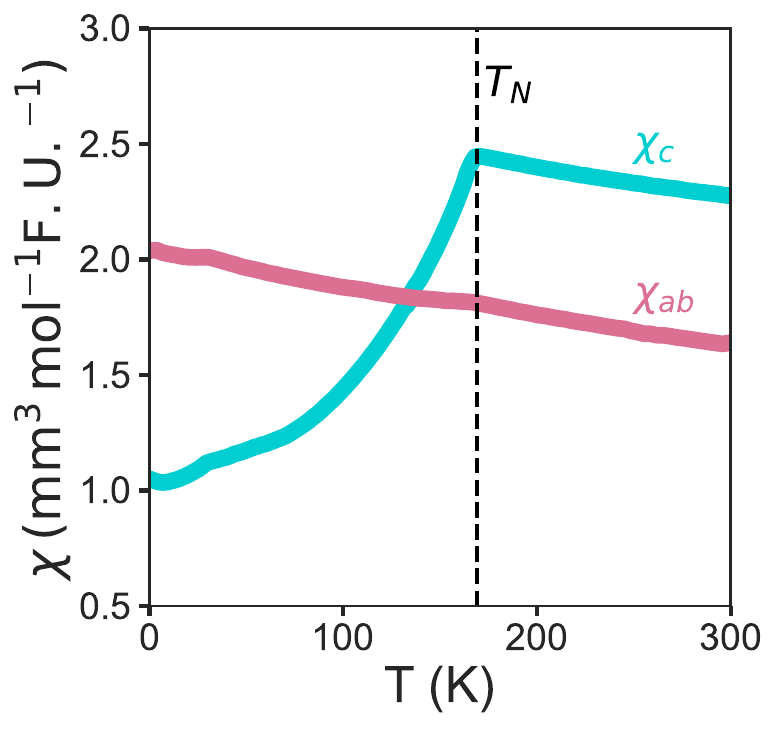}
\caption{\textbf{Collinear Antiferromagnetic Order in CoNb$_4$Se$_8$}. (A) Magnetic susceptibility measured in  a magnetic field of 0.1 T applied along c-axis ($\chi_c$) and within the ab-plane ($\chi_{ab}$), for the same sample as reported in 
Ref~\cite{regmi2024altermag_arxiv}. 
}
\label{fig:afm}
\end{figure}
Fig. \ref{fig:afm} presents the temperature-dependent magnetic susceptibility of  $\mathrm{CoNb_4Se_8}$ as a function of applied magnetic field $\vec{B}$, on the same sample as reported in ref~\cite{regmi2024altermag_arxiv} . When $\vec{B}|| c$ , the susceptibility ($\chi_c$) exhibits a sharp decrease below 168 K, whereas when $\vec{B}$  is along the $ab$-plane, the susceptibility $\chi_{ab}$ manifests a kink at 168 K followed by a gradual and modest increase at lower temperatures. This behavior is characteristic of textbook antiferromagnetic ordering with moments aligned parallel to the $c$-axis, with a Néel ordering temperature of 168 K~\cite{Blundell2001}. 

\subsubsection*{Surface Termination of CoNb$_4$Se$_8$}

\begin{figure}
\centering\includegraphics[width=0.5\textwidth]{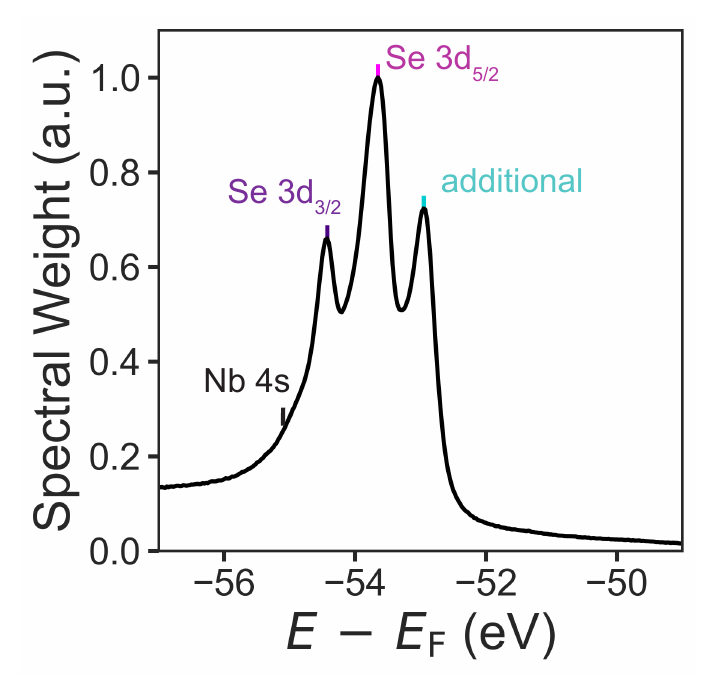}
\caption{\textbf{Xray Photoemission Spectrogram of Co-terminated CoNb$_4$Se$_8$}. XPS data taken at normal emission for a photon energy of 100 eV.}. 

\label{fig:xps}
\end{figure}
As demonstrated in ref \cite{devita2025opticalswitchinglayeredaltermagnet}, the surface termination of CoNb$_4$Se$_8$ can be determined by XPS. The Se- terminated surface reportedly has two sharp peaks around 54 eV binding energy as well as a pair of weaker peaks attributed to surface replicas\cite{devita2025opticalswitchinglayeredaltermagnet}.  By contrast, the XPS data in Fig. \ref{fig:xps} presents three distinct peaks: the two sharp peaks at $E_\mathrm{F} -54.43$ eV and $E_\mathrm{F} -53.65$ eV correspond to the Se $\mathrm{3d_{3/2}}$ and $\mathrm{3d_{5/2}}$, respectively, and the lowest energy peak at $E_\mathrm{F} -52.95$ eV, indicative of a Co-terminated surface. This additional core level could arise from the Co and Se atoms with dangling bonds at the surface, out of which form the surface valence states identified in the main manuscript. Passivation of these dangling bonds by e.g. surface adsorption of adatoms could be used to determine whether this additional core level is of surface character  \cite{Samsavar_PRB1990_surfacexps}.

\subsubsection*{Spin and Sublattice Sensitivity in ARPES Measurements}

\begin{figure}
\centering\includegraphics[width=0.8\textwidth]{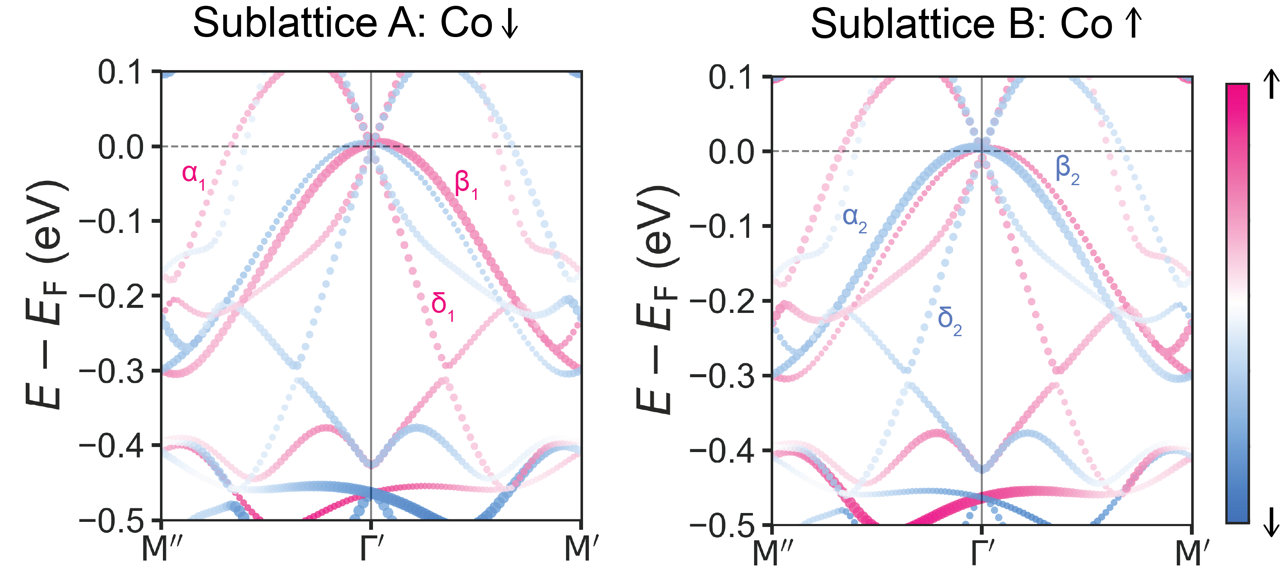}
\caption{\textbf{ARPES sublattice sensitivity in CoNb$_4$Se$_8$}. Spin-resolved DFT calculations of electronic structure along  $\mathrm{M^{\prime\prime} - \Gamma{^\prime} - M^{\prime}}$ deriving from the Co$\downarrow$ sublattice (left) and from the Co$\uparrow$ sublattice (right). Size of bands corresponds to the magnitude of their sublattice projection, whereas color denotes spin polarization along the $c$ axis.}

\label{fig:sublattice_sensitivity}
\end{figure}

In the main manuscript, the ARPES data along $\mathrm{M^{\prime\prime} - \Gamma{^\prime} - M^{\prime}}$ in Fig. \ref{fig:arpes_mgm} presents bands whose spin polarization alternates upon crossing $\Gamma^\prime$. While the DFT calculations present two pairs of spin-split bands labelled $\beta_{1,2}$ and $\delta_{1,2}$, the spin-ARPES data presents only one of each bands, $\beta$ and $\delta$ . In Fig.\ref{fig:sublattice_sensitivity} we present DFT-calculated spin- and sublattice- projected band structure in CoNb$_4$Se$_8$ along $\mathrm{M^{\prime\prime} - \Gamma{^\prime} - M^{\prime}}$. In the bands projected onto Co$\downarrow$ sublattice A (left), the $\beta_2$ band polarized spin $\downarrow$  is more prominent along $\mathrm{M^{\prime\prime} - \Gamma{^\prime}}$, whereas the  $\beta_1$ band polarized spin  $\uparrow$ is more prominent along  $\mathrm{\Gamma{^\prime} - M^{\prime}}$. Similarly, the $\delta_2$ band polarized spin $\downarrow$  is more prominent along $\mathrm{M^{\prime\prime} - \Gamma{^\prime}}$, whereas the  $\delta_1$ band polarized spin  $\uparrow$ is more prominent along  $\mathrm{\Gamma{^\prime} - M^{\prime}}$. By contrast, in the bands projected onto Co$\uparrow$ sublattice B (right), the $\beta_1, \delta_1$ bands polarized spin $\uparrow$  are more prominent along $\mathrm{M^{\prime\prime} - \Gamma{^\prime}}$, whereas the  $\beta_2, \delta_2$ band polarized spin  $\downarrow$ is more prominent along  $\mathrm{\Gamma{^\prime} - M^{\prime}}$. The observed polarization measured in ARPES: spin $\downarrow$ along $\mathrm{M^{\prime\prime} - \Gamma{^\prime}}$ and spin $\uparrow$ along $\mathrm{\Gamma{^\prime} - M^{\prime}}$,  suggests that the photoemission experiment is more sensitive to Co$\downarrow$ sublattice A than to Co$\uparrow$ sublattice B. We tentatively attribute the overall stronger spin polarization along $\mathrm{\Gamma{^\prime} - M^{\prime}}$ to a relatively larger contribution of spectral weight from NRSS bands as opposed to the the spin unpolarized narrow band at low energy.

\subsubsection*{Energy-Selective Co Superlattice Bands in CoNb$_4$Se$_8$}

\begin{figure}
\centering\includegraphics[width=0.9\textwidth]{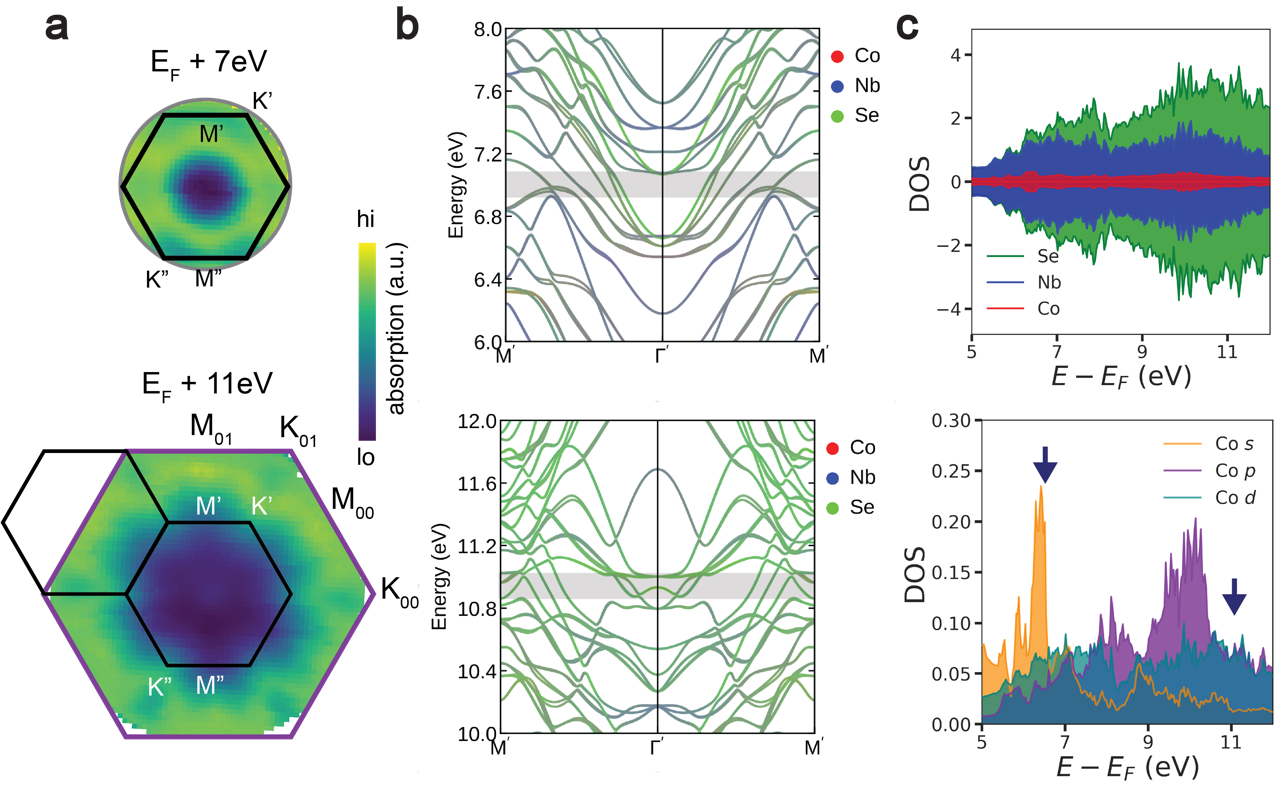}
\caption{\textbf{Energy-Selective Co Superlattice Bands in CoNb$_4$Se$_8$}. \textbf{(A)} ARRES absorption spectra at a constant energy surface of 7 eV (top) and 11 eV (bottom) above $E_\mathrm{F}$ \textbf{(B)} Atom-projected bulk DFT band structure along $\mathrm{M^{\prime\prime} - \Gamma{^\prime} - M^{\prime}}$ in energy regions corresponding to the ARRES data. \textbf{(C) }DFT-calculated atom-projected density of states}

\label{fig:superlattice_arres}
\end{figure}

Similar to other intercalated transition metal dichalcogenides, The electronic structure of CoNb$_4$Se$_8$ presents superlattice bands at energies where the intercalant orbitals hybridize with those of the host lattice  \cite{popcevic_conb3s6_arpes}. Fig. \ref{fig:superlattice_arres}A presents ARRES constant energy surfaces at 7 and 11 eV above the Fermi level; the former of which exhibits a Brillouin zone with the geometry of the 2x2 Co supercell, whereas the latter exhibits a Brillouin zone with the geometry of the parent NbSe$_2$. These findings are corroborated by atom-projected DFT calculations (Fig. \ref{fig:superlattice_arres}B) which suggest that electronic states near 11 eV are predominantly of Nb and Se character, whereas the calculated states near 7 eV host a small amount of Co character. Indeed, the atom-projected density of states (Fig. \ref{fig:superlattice_arres}C), also suggests that Co character is larger near 7 eV than at 11eV and above.

\subsubsection*{DFT and Experimental bands in the Occupied States of CoNb$_4$Se$_8$}

\begin{figure}
\centering\includegraphics[width=0.9\textwidth]{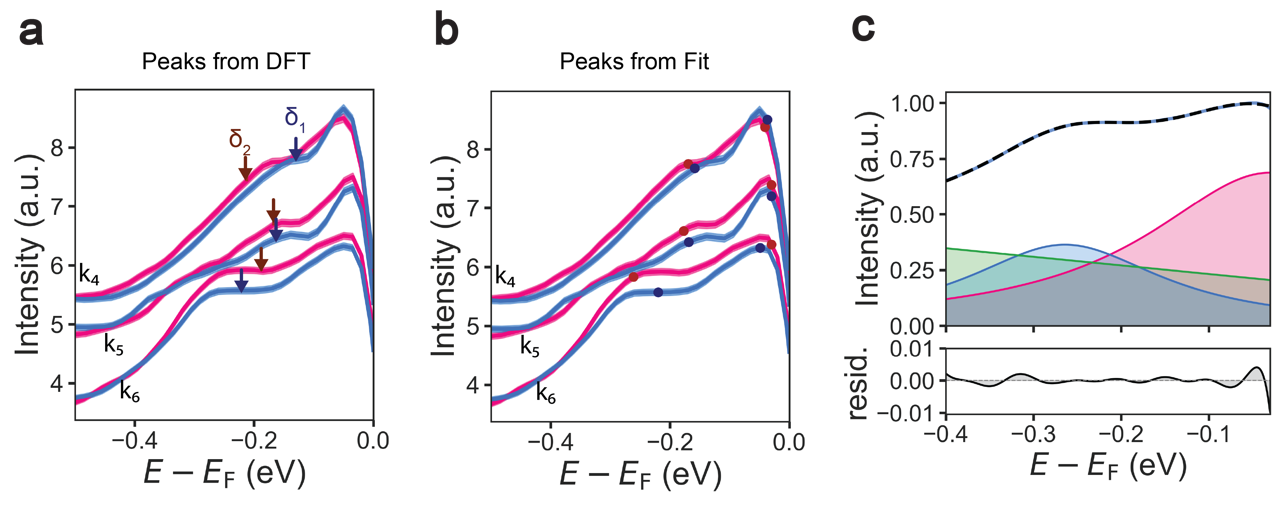}
\caption{\textbf{DFT and experimental bands in in CoNb$_4$Se$_8$}. \textbf{(A,B)} Spin-resolved EDCs spectra from Fig. \ref{fig:arpes_kgk}G, along momenta specified in Fig. \ref{fig:arpes_kgk}F in the main text. Measurement uncertainty out to to $1\sigma$ outlined in grey. Navy (red) arrows in \textbf{A} indicate DFT energies of $\delta_2$($\delta_1$) band, while Navy (red) circles in \textbf{B} indicate peaks acquired from a fit to the EDCs spectra. \textbf{(C)} Example of a Lorentzian lineshape fit to the EDCs spectra at momentum $k_6$.}

\label{fig:spin_edc_fits}
\end{figure}

To validate our interpretation of the spin-resolved EDCs data in the main manuscript, in Fig. \ref{fig:spin_edc_fits} we compare the experimental peak positions with the band energies predicted by DFT. Peak positions are extracted experimentally a pair of Lorentzian lineshapes and an affine background to data within a range where the Fermi-Dirac distribution is not present, i.e. ($E < E_\mathrm{F} - 4k_BT$ ). A representative fit at momentum $k_6$ is shown in Fig.~\ref{fig:spin_edc_fits}C. Across momenta $k_4 -k_6$ , the fitted peak energies are found to be in qualitative agreement with the DFT predictions, lying within 100 meV of the calculated band energies. This correspondence supports our assignment of the two spectral features to the spin-split $\delta_1$ and $\delta_2$ bands and provides evidence that the observed spin polarization in CoNb$_4$Se$_8$ arises from the NRSS band structure predicted by DFT. The presence of a much stronger low energy narrow band, and weaker intensity of the $\delta$ bands along momenta $k_1 -k_3$  preclude experimental assignment of peak locations. 

\end{document}